\documentclass[preprint, sigplan]{acmart}
\renewcommand\footnotetextcopyrightpermission[1]{} % removes footnote with conference information in first column

\usepackage[ruled,vlined]{algorithm2e}
\usepackage{algpseudocode}
\usepackage{balance}
\usepackage{mathrsfs}
\usepackage{microtype}
\usepackage{xspace}
\usepackage{booktabs}
\usepackage{subcaption}

\algnewcommand\algorithmicinput{\textbf{Input:}}
\algnewcommand\Input{\item[\algorithmicinput]}

\algnewcommand\algorithmicoutput{\textbf{Returns:}}
\algnewcommand\Output{\item[\algorithmicoutput]}

\newcommand{\tablesize}{}

\title{Vectorized Secure Evaluation of Decision Forests}
\newcommand{\system}{\text{COPSE}\xspace}
\newcommand{\reponame}{\url{https://bitbucket.org/plcl/copse/}}
\author{Raghav Malik}
\affiliation{
    \department{School of Electrical and Computer Engineering}
    \institution{Purdue University}
    \city{West Lafayette}
    \state{IN}
    \country{USA}}
\email{malik22@purdue.edu}
\author{Vidush Singhal}
\affiliation{
    \department{School of Electrical and Computer Engineering}
    \institution{Purdue University}
    \city{West Lafayette}
    \state{IN}
    \country{USA}}
\email{singhav@purdue.edu}
\author{Benjamin Gottfried}
\affiliation{
    \department{School of Electrical and Computer Engineering}
    \institution{Purdue University}
    \city{West Lafayette}
    \state{IN}
    \country{USA}}
\email{bg@purdue.edu}
\author{Milind Kulkarni}
\affiliation{
    \department{School of Electrical and Computer Engineering}
    \institution{Purdue University}
    \city{West Lafayette}
    \state{IN}
    \country{USA}}
\email{milind@purdue.edu}

\setcopyright{rightsretained}
\acmPrice{}
\acmDOI{10.1145/3453483.3454094}
\acmYear{2021}
\copyrightyear{2021}
\acmSubmissionID{pldi21main-p522-p}
\acmISBN{978-1-4503-8391-2/21/06}
\acmConference[PLDI '21]{Proceedings of the 42nd ACM SIGPLAN International Conference on Programming Language Design and Implementation}{June 20--25, 2021}{Virtual, Canada}
\acmBooktitle{Proceedings of the 42nd ACM SIGPLAN International Conference on Programming Language Design and Implementation (PLDI '21), June 20--25, 2021, Virtual, Canada}

\begin{document}

\begin{abstract}
    As the demand for machine learning--based inference increases in tandem with concerns about privacy, there is a growing recognition of the need for secure machine learning, in which secret models can be used to classify private data without the model or data being leaked.
    Fully Homomorphic Encryption (FHE) allows arbitrary computation to be done over encrypted data, providing an attractive approach to providing such secure inference.
    While such computation is often orders of magnitude slower than its plaintext counterpart, the ability of FHE cryptosystems to do \emph{ciphertext packing}---that is, encrypting an entire vector of plaintexts such that operations are evaluated elementwise on the vector---helps ameliorate this overhead, effectively creating a SIMD architecture where computation can be vectorized for more efficient evaluation.
    Most recent research in this area has targeted regular, easily vectorizable neural network models.
    Applying similar techniques to irregular ML models such as decision forests remains unexplored, due to their complex, hard-to-vectorize structures.

    In this paper we present \system, the first system that exploits ciphertext packing to perform decision-forest inference. \system consists of a staging compiler that automatically restructures and compiles decision forest models down to a new set of vectorizable primitives for secure inference.
    We find that \system's compiled models outperform the state of the art across a range of decision forest models, often by more than an order of magnitude, while still scaling well.
\end{abstract}

\begin{CCSXML}
    <ccs2012>

    <concept>
    <concept_id>10011007.10011006.10011041</concept_id>
    <concept_desc>Software and its engineering~Compilers</concept_desc>
    <concept_significance>500</concept_significance>
    </concept>

    <concept>
    <concept_id>10011007.10011006.10011041.10011048</concept_id>
    <concept_desc>Software and its engineering~Runtime environments</concept_desc>
    <concept_significance>500</concept_significance>
    </concept>

    <concept>
    <concept_id>10010147.10010257.10010293.10003660</concept_id>
    <concept_desc>Computing methodologies~Classification and regression trees</concept_desc>
    <concept_significance>500</concept_significance>
    </concept>

    <concept>
    <concept_id>10002978.10003022.10003028</concept_id>
    <concept_desc>Security and privacy~Domain-specific security and privacy architectures</concept_desc>
    <concept_significance>500</concept_significance>
    </concept>

    </ccs2012>
\end{CCSXML}

\ccsdesc[500]{Software and its engineering~Compilers}
\ccsdesc[500]{Software and its engineering~Runtime environments}
\ccsdesc[500]{Computing methodologies~Classification and regression trees}
\ccsdesc[500]{Security and privacy~Domain-specific security and privacy architectures}

\keywords{Homomorphic Encryption, Decision Forests, Vectorization}

\maketitle
\pagestyle{plain}
\section{Introduction}
In recent years, there has been substantial interest in secure machine learning: applications of machine learning where the ``owners'' of a model, input data, or even the computational resources may not be the same entity, and hence may not want to reveal information to one another. Settings where these applications are important include banks sharing financial data while complying with financial regulations, hospitals sharing patient data while adhering to HIPAA, and users offloading sensitive computation to cloud providers.

There are many ways of implementing secure machine learning algorithms, with different tradeoffs of efficiency and privacy.
One popular approach, thanks to its generality, is based on {\em fully homomorphic encryption} (FHE)~\cite{FHE}.  FHE is a cryptosystem that allows performing homomorphic addition and multiplication over asymmetrically encrypted ciphertexts such that when encryptions of integers are homomorphically added the resulting ciphertext is the encryption of their sum, and similarly when the ciphertexts are homorphically multiplied the result is an encryption of their product. In other words, the inputs to an addition or multiply can be encrypted, the operation can be carried out over the {\em encrypted} data, and the decrypted result will be the same as if the operation were carried out on plaintext.

FHE is attractive because it is {\em complete}: we can structure arbitrarily complex calculations as an arithmetic circuit consisting of additions and multiplications, and an entity can carry out these calculations entirely over encrypted data without ever being trusted to see the actual data. 
%Moreover, extensions have been proposed for FHE that allow multiple entities to contribute inputs to the FHE circuit, meaning that multiple parties and computational services can collaborate to evaluate arbitrary computations without revealing sensitive data to one another~\cite{chen2017BatchedMM,EfficientMKFHE,asharov}.
%
Unfortunately, homomorphic operations over ciphertexts tend to be orders of magnitude slower than their plaintext counterparts, and this problem only gets worse when the ciphertext size increases, which can happen due to a higher multiplicative depth in the arithmetic circuit or a larger security parameter (meaning more-secure encryption).
As a result, most real-world applications tend to produce FHE circuits that are impractically slow to execute.

The ability of FHE cryptosystems to do \emph{ciphertext packing} somewhat mitigates this problem.
Ciphertext packing refers to encrypting a vector of integers into a single ciphertext, so that operations over that ciphertext correspond to the same operations elementwise over the vector~\cite{packedfhe}.
If a circuit can be expressed as computing over such packed ciphertexts, the total number of homomorphic operations decreases and it often scales better to larger inputs, both of which result in a more efficient circuit for the same application.
The challenge, of course, is vectorizing arbitrary computations in this way.

Much recent work in this space has focused on securely evaluating neural network--based models. Neural nets are an attractive target for FHE because the core computations of neural nets are  additions and multiplications, and in dense, feed-forward neural nets, those computations are already naturally vectorized. Recent work developed approaches that compile simple neural net specification to optimized and vectorized FHE implementations~\cite{CHET}.

However, neural nets are not the only type of machine learning model that can benefit from the advantages of secure computation. For many applications and data sets, especially those over categorical data, {\em decision forests} are better suited to solving the classification problem than neural nets.

Unfortunately, decision forests are inherently trickier to map to vectorized FHE than neural nets. The comparisons performed at each branch in a decision tree (e.g., ``is x greater than 3?'') are harder to express using the basic addition and multiplication primitives of FHE, especially if the party providing the comparison ($x > 3$?) is different than the party providing the feature ($x$). Moreover, traditional evaluation of decision trees is sequential: ``executing'' a decision tree involves walking along a single path in a decision tree corresponding to a sequence of decisions that evaluate to true.

Recently, researchers have shown how to express the computations of a decision tree as a boolean polynomial~\cite{blindfold,Bost2014MachineLC}. These approaches parallelize decision {\em forests} (a set of decision trees) by evaluating the polynomials of each tree independently. Nevertheless, these approaches still have limited scalability, as they evaluate each decision within a single tree sequentially, and do not exploit the ciphertext-packing, SIMD capabilities of FHE.

This paper shows how a compiler can restructure decision forest evaluation to more completely parallelize their evaluation and exploit the SIMD capabilities of FHE, providing scalable, parallel, secure evaluation of decision forests.

\subsection{\system: Secure Evaluation of Decision Forests}
The primitives that a cryptosystem like FHE provides can be thought of as an instruction set with  semantics that guarantee noninterference; that is, no sensitive information can be leaked through publicly measurable outputs.
One key aspect of FHE's noninterference guarantee is that it disallows branching on secret data, instead requiring that all computations be expressed as combinatorial circuits that must be fully evaluated regardless of the input.
In particular, it guarantees resistance to timing side-channel attacks in which an attacker learns some useful information about a system (such as the sequence of decisions taken at each branch in a tree) by measuring execution time or path length on various inputs.
We propose a system called \system that leverages these semantics to relax the control flow dependences in traditional decision forest programs, allowing us to restructure the inherently sequential process of decision tree evaluation into one that maps directly into existing vectorized, efficient FHE primitives.
The vectorized evaluation strategy we present here is in contrast with the traditional polynomial-based strategy presented by \citet{blindfold}, which we discuss in more detail in Section~\ref{sec:prior}.

The restructured computation consists of four stages: a {\em comparison} step in which all the decision nodes are evaluated (in parallel), a {\em reshaping} step in which decisions are shuffled into a canonical order, a {\em level processing} step where all decisions at a particular depth of the tree are evaluated, and an {\em aggregation} step in which the results from each depth are combined into a final classification.

\system consists of two parts: a compiler, and a runtime.
The compiler translates a trained decision forest model into a C++ program containing a vector encoding the tree thresholds, and matrices that encode the branching shape.
The generated C++ links against the \system runtime, which loads the model and provides functions to encrypt it, encrypt feature vectors, and classify encrypted feature vectors using encrypted models.
The runtime uses HElib \cite{HElib}, which provides a low-level interface for encrypting and decrypting and homomorphically adding and multiplying ciphertext vectors, as well as providing basic parallelism capabilities through NTL (Number Theory Library)~\cite{ntl}.

\subsection{Summary of contributions}

This paper makes the following contributions:

\begin{itemize}
    \item A vectorizing compiler that translates decision forest models into efficient FHE operations.
    \item An analysis of the complexity of the generated FHE programs, showing that our approach produces low-depth FHE circuits (allowing them to be evaluated with relatively low overhead) that efficiently pack computations (allowing them to be vectorized effectvely).
    \item A runtime environment built on HElib that encrypts compiled models and executes secure inference queries.
\end{itemize}

We generate several synthetic microbenchmarks and show that while there is a linear relationship between level processing time and the number of decisions in the model, some of the work (such as the comparison step) is done only once and takes a constant amount of time regardless of the model size.
We also train our own models on several open source ML datasets and show that not only does our compiler generate FHE circuits that perform classification several times faster than previous work, \system can also easily scale up to much larger models by exploiting the parallelism we get ``for free'' from this restructuring.

While there are FHE schemes that allow for three distinct parties (the server, the model owner, and the data owner), either by using multi-key constructions or by using some protocol to compute a secret key shared between the data and model owners, neither option is currently implemented in HElib (as of version 1.1.0).
Therefore, we test our benchmarks on scenarios in which there are only two real parties: either the model and data owners are the same party offloading computation, or the server also owns the model and allows the data owner to use it for inference.

%\subsection{Outline}
%
%The remainder of this paper is structured as follows.
%Section~\ref{sec:background} provides background on decision forests, fully-homo\-morphic encryption, and related work in both the secure evaluation of decision forests as well as compiler optimization of (insecure) decision forest evaluation.
%Section~\ref{sec:overview} gives an overview of the vectorized algorithm,
%Section~\ref{sec:algorithm} discusses the details of the algorithm, along with the notation and primitives needed to formalize it.
%Section~\ref{sec:compiler} gives details on the design and implementation of the \system compiler and runtime system.
%Section~\ref{sec:complexity} characterizes the complexity of the evaluation algorithm.
%Section~\ref{sec:parties} discusses the different configurations for the physical parties involved in the inference algorithm
%Section~\ref{sec:evaluation} evaluates \system on microbenchmarks as well as real decision forest models, and Section~\ref{sec:conclusions} concludes.
\section{Background}\label{sec:background}
\subsection{Decision Forests}
\begin{figure}
    \centering
    \includegraphics[width=0.8\columnwidth]{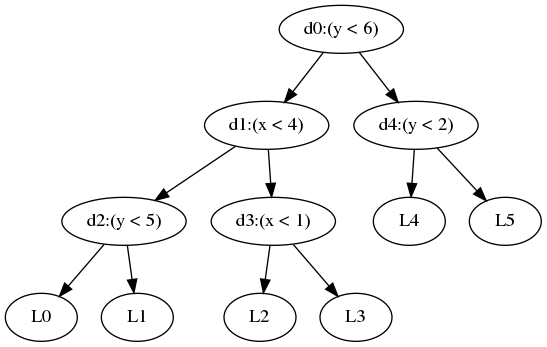}
    \vspace{-1em}
    \caption{Example decision tree}
    \Description{Example decision tree with 5 decision nodes and a maximum level of 3}
    \label{fig:example-tree}
\end{figure}
A \emph{decision tree} is a classification model that assigns a class label to a vector of features by sequentially comparing the features against various thresholds.
Figure~\ref{fig:example-tree} shows an example of a single decision tree.
Inference over a decision tree is recursive.
Leaf nodes of the tree correspond to class labels, whereas each interior ``branch'' node specifies a feature and a threshold.
That feature from the vector is compared against the threshold, and depending on the result of the comparison either the left or right child of the tree is evaluated.
For instance, the tree in Figure~\ref{fig:example-tree} uses $x$ and $y$ as its features and assigns class labels $L_0$ -- $L_5$.
Assuming the left branch is taken when the decision is false and the right branch is taken when true, the tree assigns the input feature $(x, y) = (0, 5)$ to the class label $L_4$.
A \emph{decision forest} model consists of several decision trees over the same feature set in parallel.
Inference over a decision forest usually consists of obtaining a class label from each individual tree, and then combining the labels in some way (either by averaging or choosing the label selected by the plurality of trees, or some other domain-specific combining function).
\subsubsection{Vectorizing}
Making inferences over a decision forest is an inherently sequential process.
Each branch must be evaluated to determine which of its children to evaluate next; these serial dependences make it difficult to vectorize the evaluation of a single tree.
Such dependences do not exist within individual trees of a forest; However, the nonuniform branching and depth of each tree makes them unsuitable for directly packing into vectors for evaluation. \cite{CGO.2013.6494989}
% \subsubsection{Preserving Security}\label{sec:preserving-security}
% In any system that carries out privacy-preserving computation, noninterference is an important property to maintain.
% That is, we have to ensure not only that the encryption schemes and protocols being used are secure against various adversarial models, but also that the design of the system does not unwittingly leak sensitive or protected information.

% A privacy-preserving system must be designed robustly against side-channel attacks; in particular, timing-based side channel attacks, where sensitive information can be learned by observing the time various parts of the protocol take to execute.
% Practically speaking, this puts a burden on how trees are evaluated.
% The protocol for evaluating a decision tree in plaintext --- i.e., evaluating the root node, choosing either the left or right node based on the result, and recursively descending until a label is reached --- is no longer a viable option.
% When the tree is unbalanced, different paths from the root to labels may have different lengths.
% As each path is executed sequentially, an attacker can learn information about the length of the path by studying the time it takes to evaluate.
% Furthermore, as the path taken is determined uniquely by the inputs, this means that an attacker can potentially learn information about the nature of the inputs to the model simply by studying its run time.
% To guard against this, we need to ``pad out'' the execution, and evaluate every path through the tree no matter what the input features are.

\subsection{Fully Homomorphic Encryption}
Fully Homomorphic Encryption, or FHE, is an asymmetric encryption scheme which allows computation to be carried out over ciphertexts that are encrypted under the same public key without needing to know the private decryption key.\footnote{Note that {\em fully} homomorphic encryption is distinguished from {\em partially} homomorphic encryption in that the latter only supports either addition or multiplication, while the former supports both.} \cite{FHE} In other words, an {\em evaluator} can perform computation on {\em encrypted} data to produce an encrypted version of the result {\em without ever seeing the plaintext data}. Decrypting this encrypted result will yield the same result as if the entire computation was performed in plaintext. This makes FHE an attractive option for users to offload computation to servers without worrying that the computational server will get access to sensitive data.
%\ben{Does it need to be emphasized that the results of FH operations on ciphertext, when decrpyted, are identical to the same ones performed on the plaintext?}

\subsubsection{Limitations}
One major limitation of FHE cryptosystems is the speed.
Homomorphic additions and multiplications over ciphertexts are orders of magnitude slower than equivalent operations over plaintext \cite{CHET}.
This makes it difficult to directly translate large computations into an arithmetic circuit to be evaluated in FHE, because such circuits often take prohibitively long to execute.

Another limitation we have to deal with in FHE systems is that every multiplication introduces some noise into the ciphertext which, when sufficiently accumulated, makes it impossible to decrypt. Hence, a homomorphic circuit needs to be designed with {\em multiplicative depth}---the maximum number of multiplications in a dependence chain---in mind to ensure that the computation result can be recovered.
This is a problem partially solved by ``bootstrapping'', which involves homomorphically reencrypting a ciphertext to remove the noise \cite{FHE}.
Bootstrapping is not a perfect solution, however, as it is an expensive operation that takes a lot of time.
We can attempt to support a higher multiplicative depth in the circuit before bootstrapping is necessary by increasing the security parameter\footnote{The security parameter is a parameter of the encryption scheme that specifies how many ``bits of security'' we get by encrypting a ciphertext. In general, increasing this parameter makes the encryption stronger (harder to break, a higher maximum multiplicative depth) at the cost of making computation more expensive.}, however this results in larger ciphertexts which are slower and more expensive to compute over.
Thus, when expressing any reasonably-sized computation in FHE, we have to optimize against having a high multiplicative depth.

\subsubsection{Advantages}\label{sec:fhe-advantages}
FHE schemes create opportunities for securely computing functions of secret inputs between distrusting parties, as well as for offloading the processing of secure data to an untrusted server.
This is useful in a machine learning domain, as it is easy to imagine use cases where one party has trained a model, and they wish to allow others to make inferences over it without revealing the structure or details of the model itself (in the case of decision forests, without revealing the thresholds).
Alternatively, we see this being applicable in a setting where one party has sensitive data they wish to classify using a third-party model.
%\ben{Emphasize this aspect because it's this use-case that makes the overhead for FHE worth it rather than decrypting prior to evaluation.}
In the most general scenario, and perhaps the one with widest applicability, we have one party with sensitive data and another with a sensitive model, both of which can be offloaded to an untrusted third party server for inference without revealing details about the data or the model to the other party.

The feature of FHE schemes that we will make the most use of is the notion of \emph{ciphertext packing}.\cite{Brakerski12packedciphertexts}
This refers to encrypting an entire vector of plaintext values into a single ciphertext in such a way that homomorphic additions and multiplications over the ciphertext correspond to elementwise additions and multiplications over the plaintext vector.
This effectively gives us a SIMD (single instruction multiple data) architecture to target, where the vector widths are much larger than they typically are for physical SIMD architectures.
This means that if we are careful about our data representation, we can leverage the ability to pack data into vectors in order to mitigate the high runtime costs of working with encrypted data.

%\raghav{I kind of just word-vomited here. I think I've expressed the right ideas, but I'm not convinced there isn't a better way to express them.}
\subsubsection{Non-interference}
Fundamental to the secure computation is {\em non-interference}. The contents of private data should not ``leak'' and produce outputs (data or behavior) observable by other parties. One vector of leakage is through conditional execution: if the result of a branch is dependent on private data, an observer may gain information by observing the resulting execution of the program (e.g., if different paths through the program take different amounts of time).

A common method to prevent this leakage, enforced by approaches such as FHE, is branchless programming. Rather than evaluating a conditional and taking a branch, {\em both} paths of a conditional expression are evaluated, and some homomorphic computation is performed to produce the desired result (for instance, multiplying the return values of each path by a boolean ciphertext to select the right result).

This execution strategy seems like it presents a problem for decision forests. The standard sequential algorithm for evaluating a decision tree inherently evaluates a number of conditional branches to select the final label. The FHE evaluation strategy effectively requires us to ``pad out'' the execution by evaluating every branch of the tree, and only selecting the final label at the end. The key insight of our paper is that this seeming limitation affords an opportunity: parallel evaluation.

\subsection{Related Work}
Related work in this area falls broadly into three categories: making the inference process over decision forest models secure, vectorizing the evaluation of such models, and vectorizing secure inference over general machine learning models.

\subsubsection{Securely evaluating decision trees}
%\raghav{Is that enough?} 
The two main approaches to securely performing inference on decision forest models are oblivious transfer (OT) based methods such as the one found in \citet{PrivatelyEvaluatingDecisionTreesandRandomForests}, and constructing polynomial representations of the trees as seen in \citet{blindfold} and \citet{Bost2014MachineLC}.
The OT approach involves using rounds of oblivious transfer to allow the client to interactively select a path through the forest without revealing details about this path to the evaluator.
The polynomial-based methods represent each tree as a boolean polynomial, where each decision node is a variable, each label node corresponds to a term in the polynomial, and the boolean decisions multiplied together in each term encode the path from the root of the tree to the label.

% There has been prior work in the area of securely evaluating decision trees and decision forests using FHE \cite{blindfold,Bost2014MachineLC}, however most of this work either does not address vectorizability, or does not take full advantage of the SIMD parallelism offered by FHE schemes.

\citet{PrivatelyEvaluatingDecisionTreesandRandomForests} use additive homomorphism to interactively compute each decision result between the server and the client.
The client then decrypts these decision results and uses them as the input to a round of oblivious transfer (OT) with the server, which results in the client learning only the final class label, and the server not learning anything about the decision results.
To hide the tree structure from the client, the server first adds dummy nodes to the tree and then randomly permutes the branches.
This evaluation protocol relies on the model being available in plaintext to the server, which is a restriction we overcome by providing a way to represent the model as a series of ciphertexts, allowing its structure to be hidden from both the client and the server.

%\milind{I don't know what ``boolean polynomials in the comparison results'' means. I think we should expand this a bit with, e.g., a small example: ``for example, a decision tree with a single node would look like ...''}\raghav{Better?}
\citet{Bost2014MachineLC} and \citet{blindfold} structure the tree as a vector of boolean polynomials in the comparison results, each returning a single bit of the class label.
For example,  for a decision tree with a single branch $d_0$, and $L_0$ and $L_1$ as the true and false labels respectively, the $i^{th}$ polynomial would look like $p_i(d_0) = d_0L_0^i + (1 - d_0)L_1^i$ (where $L_j^i$ denotes the $i^{th}$ bit of the $j^{th}$ label).
The multiplications in each term are evaluated recursively in pairs to give a multiplicative depth that is logarithmic in the order of the polynomial instead of linear.
Since the polynomials for each bit are over the same decision results, they are packed into SIMD slots so that each SIMD operation works over all the bits in parallel.
There is no SIMD capability beyond this, as every decision node in the tree is still evaluated sequentially.
This is different from our technique, which exploits SIMD parallelism between unrelated sets of decision nodes.
Since the number of decision nodes in a forest is roughly exponential in the number of bits in the class labels, we expect our technique to scale better to larger models.

\subsubsection{Evaluating vectorized decision forests}
Some work has been done in the area of vectorizing the evaluation of decision forests.
\citet{CGO.2013.6494989} lay out each tree regularly in contiguous memory to turn the control dependence of each branch into a data dependence.
They propose a protocol to evaluate a vector of decision nodes and produce a new vector of the results.
Since there are no control dependencies anymore, each tree in the forest can be packed into a single SIMD slot, allowing for vecotorized evaluation of the entire forest.
While this method results in evaluating the entire forest top-down without much extra work, it does not directly work for our case.
At each step, all the current nodes are evaluated and produce the index in continguous memory of the next node to evaluate.
This requires random-access to the memory where the nodes of the tree are stored, which is not possible to implement efficiently in an FHE setting.

\subsubsection{Vectorized inference}
\citet{CHET} propose CHET, a compiler for homomorphic tensor programs.
CHET analyzes input tensor programs such as neural network inference, and determines an optimal set of encryption parameters, as well as the most efficient data layout for easily vectorizing the computation.
Although CHET improves performance with regular data structures like tensors, it is not build to deal with fundamentally irregular programs like decision forests.
The \system compiler handles this irregularity by severing the dependences within each tree and producing easily vectorized structures.
\label{sec:prior}
% describe what the NDSS paper does
% describe what the Privately Evaluating Decision Trees and Random Forests paper does
% describe what the https://eprint.iacr.org/2019/1282.pdf paper does
% describe what the blindfold paper does

\section{Overview}\label{sec:overview}
This section gives a high level overview of the vectorizable decision forest evaluation algorithm before diving into the details.
We will use the decision tree in Figure~\ref{fig:example-tree} as a running example to illustrate these steps.

\subsection{The Players}
The algorithm deals with three abstract entities: the model owner (Maurice), the data owner (Diane), and the server that performs the computation (Sally).
These could correspond to different physical parties who want to conceal information from each other, or multiple entities could map to the same physical party (for example, the same party could own the model and the server).
Section~\ref{sec:parties} discusses the security implications of different configurations.
The three entities each own different components of the system, and work together to evaluate a decision forest for a set of features.
\begin{enumerate}
    \item Maurice owns a decision forest model for which the features and labels are public, but he wants to keep the shapes of the trees and their threshold values secret
    \item Diane owns several feature vectors that she wants to classify using the model owned by Maurice
    \item Sally owns no data but possesses computational power and allows Maurice and Diane to offload their computations to her.
\end{enumerate}

\subsection{The Workflow}

\begin{figure}[tb]
    \centering
    \includegraphics[width=\columnwidth]{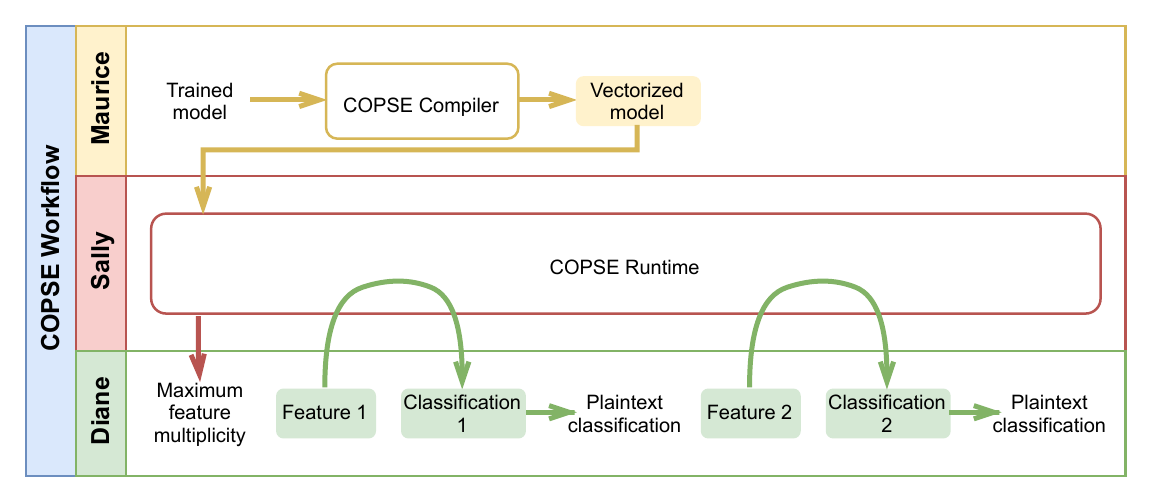}
    \caption{High-level COPSE system workflow. Yellow components and data are Maurice's responsibility. Red components are Sally's. Green components are Diane's. Shaded boxes represent encrypted data.}\label{fig:workflow}
    \Description{A three-row workflow chart of the \system system. The first column shows the model owner Maurice encrypting a trained model and sending it to the evaluator Sally in the middle column. The third column shows the data owner Diane encrypting her features to send to Sally, who responds with an encrypted classification.}
\end{figure}

Figure~\ref{fig:workflow} shows a high level overview of the COPSE workflow.
The \system system consists of two main components: a compiler used by Maurice, and a runtime used by Sally.

Once Maurice has trained a decision forest model, he uses the \system compiler to generate an encrypted and vectorized representation that he can send to Sally, who can then accept inference queries, and use the \system evaluation algorithm (summarized next) to make classifications.
When Diane wants to make a query, she first encrypts her features and then sends them to Sally, who uses the \system runtime to classify them against Maurice's encrypted model.
Sally then sends the encrypted classification result back to Diane, who can then decrypt it and send additional inference queries to the model if she wishes.

\subsection{The Evaluation Algorithm}
The multi-party evaluation algorithm proceeds in the following steps:
\paragraph{Step 0: Features} First, Maurice reveals the maximum multiplicity of any feature in a tree of the model to Sally, who then reveals it to Diane to enable the latter to set up an inference query.
In the example in Figure~\ref{fig:example-tree}, this is $3$, which corresponds to the feature $y$, as it shows up in $d_0$, $d_2$, and $d_4$, whereas $x$ only shows up in $d_1$ and $d_3$ (and hence has a multiplicity of 2).
Diane then replicates each feature in her feature vector a number of times equal to this maximum multiplicity (for instance, yielding $[x, x, x, y, y, y]$), encrypts the replicated vector, and sends it to Sally.

\begin{figure}
    \centering
    \begin{subfigure}{0.7\columnwidth}
        \centering
        \includegraphics[width=0.95\textwidth]{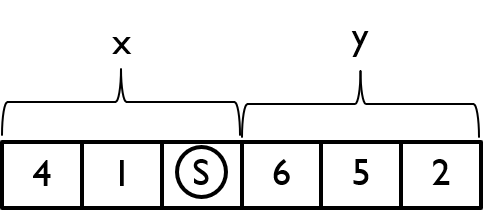}
        \caption{Padded threshold vector from annotated decision tree}\label{fig:threshold-vector}
        \Description{Visualization of annotated decision tree and padded threshold vector}
    \end{subfigure}
    \begin{subfigure}{0.95\columnwidth}
        \includegraphics[width=\textwidth]{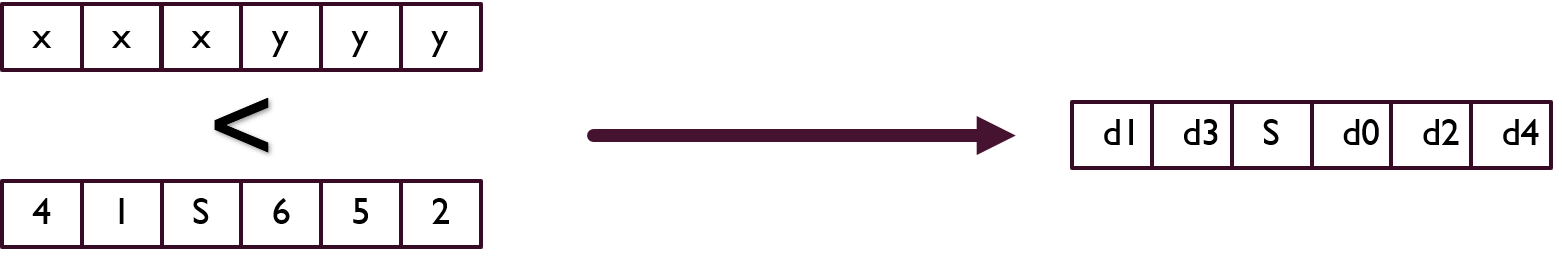}
        \caption{Comparing replicated feature vector to padded threshold vector}\label{fig:threshold-compare}
    \end{subfigure}
    \vspace{-1em}
    \caption{Illustration of vectorized comparison step}
\end{figure}

\paragraph{Step 1: Comparison} Maurice uses the \system compiler to construct a vector containing all the thresholds in the tree, grouping together thresholds from decision nodes that use the same feature.
This threshold vector is padded with a sentinel value $S$, as shown in Figure~\ref{fig:threshold-vector}, for when a feature has less than the maximum multiplicity, so that the threshold vector, representing the decision nodes in the forest, and Diane's feature vector are in one-to-one correspondence.
Once this threshold vector is constructed, Maurice sends it to Sally. To perform inference, Sally pairwise compares the vector to Diane's encrypted feature vector to get the results of every threshold comparison, as illustrated in Figure~\ref{fig:threshold-compare}.
Each element in the decision result vector is either a sentinel or corresponds to one of the decision nodes in the original tree.
Note that this thresholding happens in a single parallel step regardless of the number of branches in the tree, as both Diane's features and the decision tree thresholds are packed into vectors.

\paragraph{Step 2: Reordering}
Once Sally produces the decision vector, as in Figure~\ref{fig:threshold-compare}, she reorders all the branch decisions so they correspond to a pre-order walk of the tree, and removes the sentinel values.

\begin{figure}
    \centering
    \begin{subfigure}{0.9\columnwidth}
        \includegraphics[width=\textwidth]{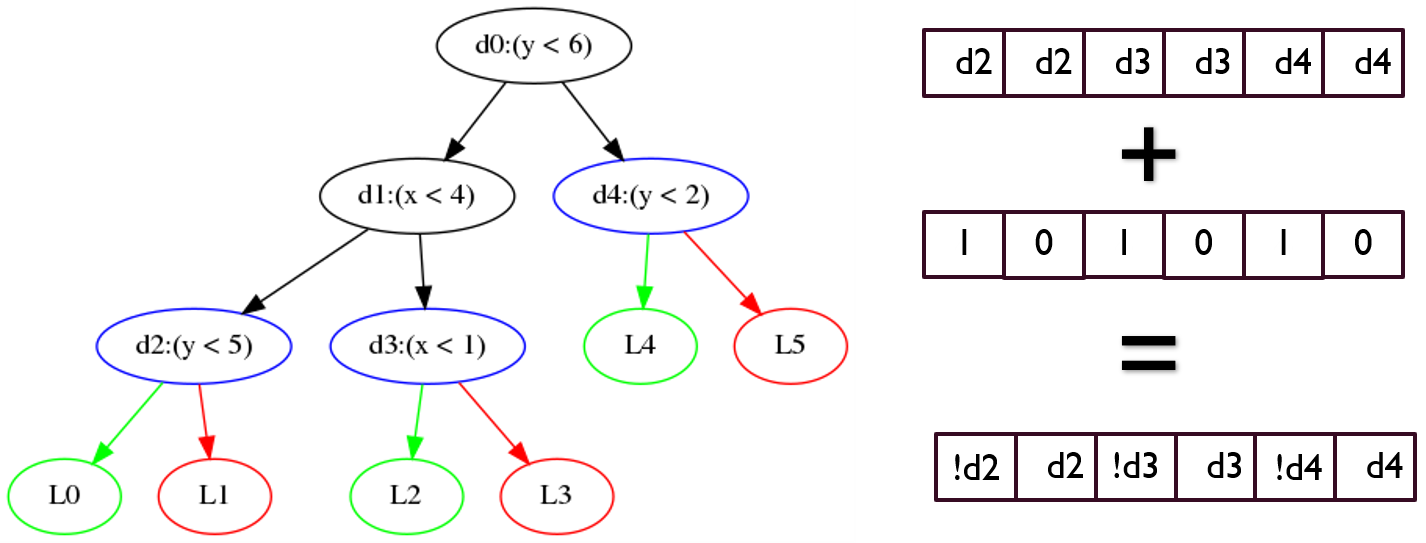}
        \caption{Processing at level 1}
    \end{subfigure}
    \begin{subfigure}{0.9\columnwidth}
        \includegraphics[width=\textwidth]{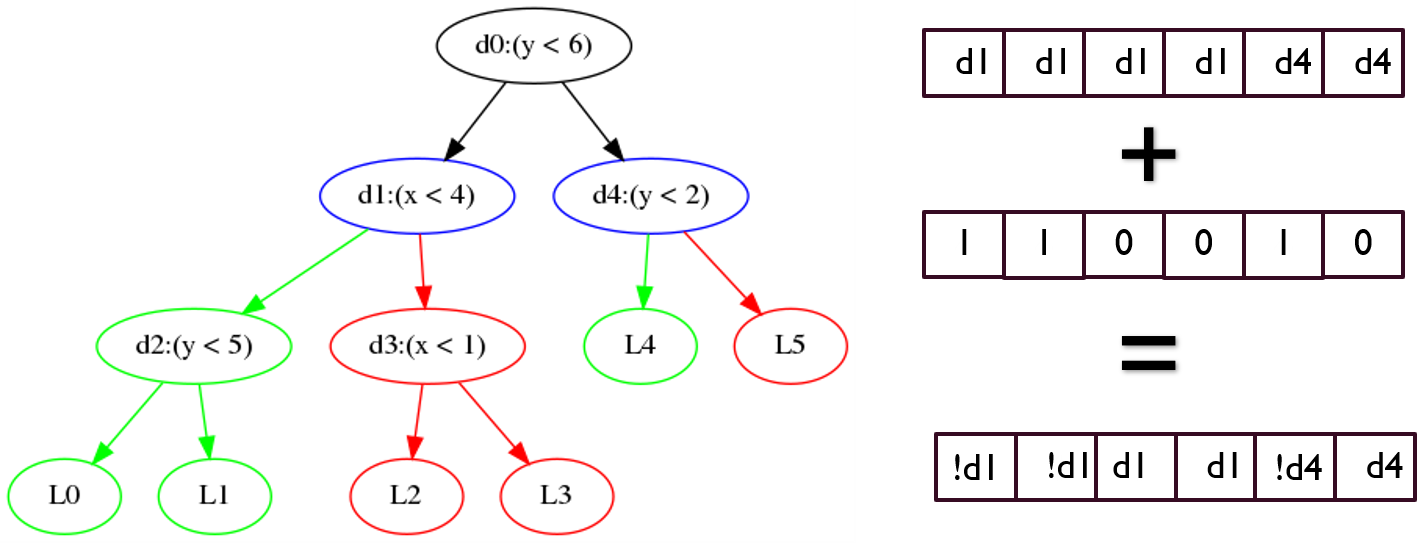}
        \caption{Processing at level 2}
    \end{subfigure}
    \vspace{-1em}
    \caption{Each level is processed individually}\label{fig:depth-masks}
    \Description{Individual level processing labels and masks}
\end{figure}

\paragraph{Step 3: Level Processing}
Now that the decision results are in a canonical order, each level (counting up from the leaves) can be processed separately.
At each level of the tree, Sally construct a binary mask encoding for each label whether it is downstream of the true or false branch at that level.
Figure~\ref{fig:depth-masks} shows how Sally processes levels 1 and 2. The nodes colored in blue are the nodes at that level, while the labels colored in green correspond to a $1$ in the mask and are downstream of the ``false'' branch, and those in red correspond to a $0$ in the mask and are downstream of the ``true'' branch. Note that $d_4$ is treated as part of level 1 {\em and} 2. This is because labels $L_4$ and $L_5$ are shallower than the other labels. Hence, $d_4$ is treated as though it exists at its own level as well as all lower levels.

%The colored labels are shown on the left in Figure~\ref{fig:depth-masks} for levels 1 and 2, with the corresponding boolean masks shown on the right.
%In the example, the decision nodes at depth 1 are $d_2$, $d_3$, and $d_4$.
%Note that $d_4$ shows up in both levels 1 and 2; this is because there is no level 2 node above $L_4$ and $L_5$, so we instead pick the level 1 node we already processed. \raghav{Does this make sense?}
Consider Level 1. Labels $L_0$, $L_2$, and $L_4$ correspond to the `false' branch for these decisions, and labels $L_1$, $L_3$, and $L_5$ correspond to the `true' branch.
The decision results corresponding to that level are picked out of the vector from step 2 and XOR'ed with the mask to yield a boolean vector encoding whether each label could possibly be the classification result given the decision results at that level (in other words, for the $i^{th}$ label to be possible, the $i^{th}$ entry in the XOR'ed vector must be true).

\begin{figure}
    \centering
    \includegraphics[width=0.9\columnwidth]{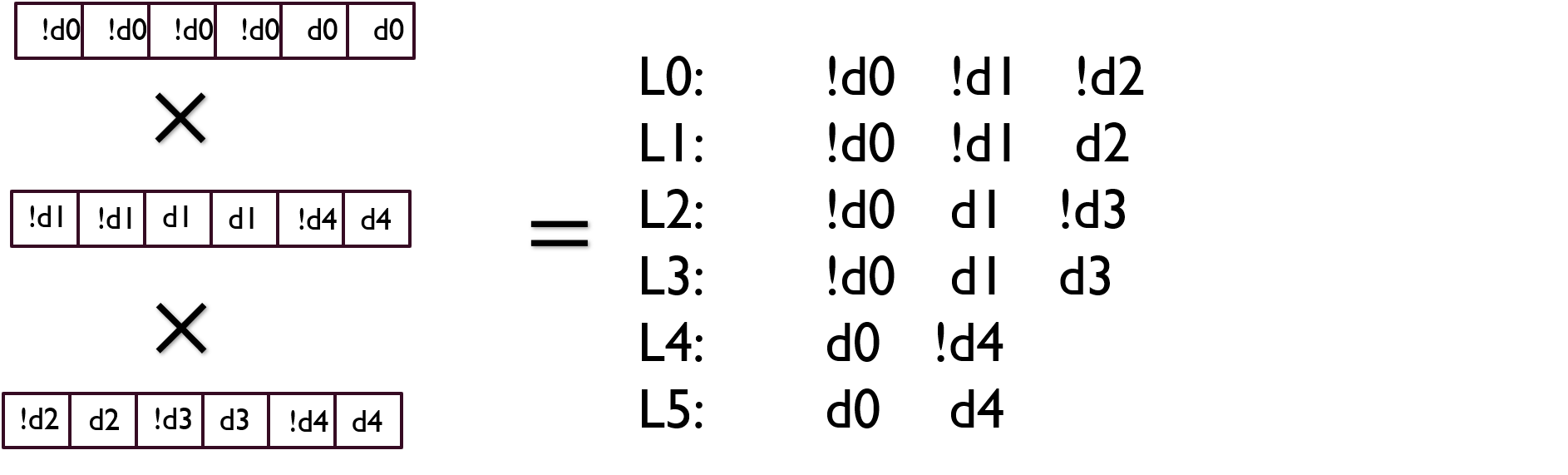}
    \vspace{-1em}
    \caption{Level vectors are multiplied to yield the final result}\label{fig:accumulation}
    \Description{Combining all level vector results to yield classification}
\end{figure}

\paragraph{Step 4: Accumulation}
Finally, once these vectors are collected for each level, Sally simply multiplies them all together to get a final label mask.
As illustrated in Figure~\ref{fig:accumulation}, an entry in this final vector is true if all the corresponding entries from the XOR'ed level vectors were true; this can only be true when the corresponding label is, in fact, the result returned by the decision tree.
Note that the return value of the evaluation algorithm is not a single label but rather an $N$-hot bitvector with a single bit turned on for each tree.
Section~\ref{sec:algorithm} discusses the reasoning behind this design choice, and Section~\ref{sec:app-security} addresses the privacy implications.

A key point to notice about this algorithm is its inherent parallelizability, as each level can be processed entirely independently of the others.
Another advantage is that all the computation at a given level can be packed into vectorized operations, as discussed in Section~\ref{sec:fhe-advantages}, which exposes even more algorithm-level parallelism.
Finally, since all these steps are performed on encrypted data using FHE, nothing about the model is revealed to Diane or Sally aside from the maximum multiplicity, and nothing about the Diane's feature vector is revealed to anybody.
Section~\ref{sec:algorithm} formalizes and describes in greater detail the exact primitives used to carry out this algorithm, and Section~\ref{sec:parties} informally discusses how configuring the FHE primitives in different ways yields different security properties.

\section{Vectorizable Evaluation Algorithm}\label{sec:algorithm}
\subsection{Preliminaries}\label{sec:preliminaries}
\subsubsection{Definitions and Important Properties}\label{sec:definitions}
\paragraph{Decision Forest:}
Consider a decision forest model $M$ consisting of trees $T_1,\dots,T_N$.
Each tree $T_i$ consists of a set of branches $B_i$ (the interior nodes) and a sequence of labels $L_i$ (the leaves).
The labels in the sequence do not necessary have to be unique.
We index all the branches in a tree by enumerating them in preorder; this indexing can be easily extended to the entire forest by not starting the count over for each new tree.
The labels of the forest are similarly (separately) indexed.

All the data of a decision forest except for its branching structure is encoded in three vectors: $\mathbf{x}$, $\mathbf{f}$, and $\mathbf{t}$.
Let $\mathbf{x} = (x_1, \dots, x_n)$ be the set of features. Then $\mathbf{f}$ is the vector encoding which feature is compared against at each branch, and $\mathbf{t}$ is the threshold at each branch.
For instance, if the branch $B_7$ has the condition $x_3 < 100$, then $f_7 = 3$ and $t_7 = 100$.
\paragraph{Properties of nodes:} Each node in the tree has a {\em level}, which is the number of branches on the longest path from the node to a label (including itself; the level of a label node is 0), a {\em downstream set}, which is the set of all labels reachable from this node, and a {\em width} which is the size of the downstream set.
An important consequence of these definitions is that, given a level $d$ and a label $L_i$, there is a unique branch node $B_j$ at level $d$ that has $L_i$ in its downstream set.
To see why this is the case, consider two distinct nodes $B_j$ and $B_k$ that contain $L_i$ in their downstream set.
One of the two must be an ancestor of the other, since each node has a unique parent; thus, they cannot have the same level.

\paragraph{Properties of models:}
We define the {\em multiplicity} $\kappa_i$ of a feature $x_i$ to be the total number of times it appears in the model (in other words, $\kappa_i$ is the number of times $i$ appears in the vector $\mathbf{f}$).
In the example tree in Figure~\ref{fig:example-tree}, $\kappa_x = 2$ and $\kappa_y = 3$.
The {\em maximum multiplicity} $K$ of a forest is the maximum multiplicity of all its features (for the example tree, $K = 3$).

The {\em branching} $b$ of a model is the total number of branch nodes it has; this is equivalent to the sum of the multiplicities for each feature, which in the example is $b = \kappa_x + \kappa_y = 2 + 3 = 5$.
The {\em quantized branching} $q$ is the product of the $K$ and the total number of features; in other words, it is the branching if every feature had maximum multiplicity.
In the example, since $K = 3$ and there are two features, $q = 6$.

\subsubsection{Data Representation and Key Kernels}\label{sec:data-rep-and-kernels}
\paragraph{Representing Non-integral Values} Rather than try to securely perform bit operations on floating point numbers, we instead represent decision thresholds as fixed-point values with the precision $p$ known at compile-time.
A vector of $k$ fixed-point values with precision $p$ is represented with $p$ bitvectors each of length $k$, with vector $i$ holding the $i^{th}$ bit of each element of the original vector. This peculiar ``transposed'' representation makes vectorizing computations easier later, allowing us to treat each bit independently while still performing comparisons in parallel.
\paragraph{Integer Comparison} We use the SecComp algorithm described by \citet{blindfold}.
Each ``bit'' of the values being compared is actually a bitvector packed as described above.
The SecComp algorithm compares two equal-length bitstrings $x$ and $y$ lexicographically.

\paragraph{Matrix Representation}
Matrices are represented as vectors of {\em generalized diagonals}.
The $i^{th}$ generalized diagonal $d_i$ of an $m\times n$ matrix $A$ is a vector defined as follows:
\[d_i = (A_{0, i}, A_{1, i+1}, \dots, A_{n-i, n}, A_{n-i+1,0}, \dots, A_{m,(m+i)\text{mod}\,n})\]
Intuitively, this is the diagonal with an offset of $i$ columns, wrapping around to the first column when necessary.
For an $m\times n$ matrix there are always $n$ generalized diagonals, each of which has length $m$.
\paragraph{Matrix Multiplication}
The diagonal representation described above makes matrix/vector multiplication easier.
To multipliy an $m\times n$ matrix $M$ by an $n\times 1$ vector $v$, we use the algorithm described by \citet{helib-algos}.
The $i^{th}$ diagonal of $M$ is multiplied component-wise by the vector $v$ rotated $i$ slots.
When $m\not\eq n$, the width of these two vectors will not be the same.
If $m > n$, $v$ is cyclically extended (e.g. $[x, y, z]$ becomes $[x, y, z, x, y, z, \dots]$).
If $n > m$, $v$ is truncated after rotating.
The vectors resulting from each such product are summed.
This has the advantage of having a constant multiplicative depth of $1$ regardless of the size of the matrix or vector.
\paragraph{Classification Result}

The classification result is encoded into a bitvector with one slot for every label node in the forest.
A slot in the bitvector holds a $1$ if the corresponding label was the one chosen by its tree, and a $0$ otherwise; in a forest with $N$ trees, $N$ slots in the bitvector will be set to $1$.

%\raghav{Is this the right place to put it?}\milind{I think we could instead move the detailed discussion to the privacy section, and just say here ``Note that this approach to generating the results yields the classification decision of {\em each} component decision tree, rather than just the plurality classification. \system chooses the former approach as one point in the tradeoff space between efficiency and privacy, as discussed in Section blah blah.''}
Note that this approach to generating the results yields the classification decision of {\em each} component decision tree, rather than just the plurality classification. \system chooses the former approach as one point in the tradeoff space between efficiency and privacy, as discussed in Section~\ref{sec:app-security-label-format}.

\subsection{Algorithmic Primitives}\label{sec:primitives}
\subsubsection{Padded Threshold Vector}
To carry out the comparisons in parallel, all the decision thresholds in the forest need to be packed into a single vector that is in one-to-one correspondence with Diane's feature vector.
This packed threshold vector is actually a sequence of $p$ bitvectors packed according to the description in Section~\ref{sec:data-rep-and-kernels}, where $p$ is the chosen fixedpoint precision (the $i^{th}$ bitvector contains the $i^{th}$ bit of each threshold).
To prevent Diane from learning the exact structure of the decisions (i.e. which feature is thresholded against at each node of the forest), we group the thresholds in the vector by the feature they correspond to (so all the $x_1$'s go at the beginning, followed by the $x_2$'s, and so on).
Revealing some information about how many times each feature is in the forest (in other words, $\kappa_i$) is, of course, unavoidable.
We limit the scope of this information leak by only revealing the maximum multiplicity $K$ of all the features; for any feature with fewer than $K$ occurences, the threshold vector is padded with some sentinel value $S$ until its effective multiplicity is $K$.
Our implementation chooses $S = 0$, but the exact value does not matter as the results from comparisons against a sentinel are removed later anyway. %\raghav{Does this paragraph make sense? I tried to avoid all the jargon from the previous iteration}

\subsubsection{Reshuffling Matrix}
Once a boolean vector is produced containing the decision result for each node of the forest, it must be rearranged to correspond to the order of the branch enumeration.
This also means removing the slots in the vector resulting from comparing against one of the sentinels used to pad the thresholds.
In order to encode this reshuffling and sentinel removal, we construct a binary matrix $\mathscr{R}$ that, when multiplied by the decision result vector, produces a new vector with the results sorted correctly.
The matrix $\mathscr{R}$ has a $1$ in row $i$ and column $j$ if the $j^{th}$ element of the padded threshold vector corresponds to the $i^{th}$ branch of the decision tree.
This means that there is exactly one of these in every row of $\mathscr{R}$, and at most one in every column, with the empty columns corresponding to the indices of the sentinel values.

\subsubsection{Level Matrices}
A level matrix is constructed for each level of the forest up to its maximum depth.
For each label, the matrix at a given level selects the branch node above the label at that level.
In the case where there is no such branch (for instance, there are branches above $L_4$ at level 1 and level 3, but none at level 2 in the example in Figure~\ref{fig:example-tree}), the highest branch not exceeding that level is selected (this is $d_4$).
The decision to do this is somewhat arbitrary; we could have just as easily chosen to use a higher level branch (such as $d_0$) instead, since what really matters is that every branch is represented in at least one of the levels.
The level matrices are, like the reshuffling matrix, boolean matrices.
A level matrix has a $1$ in row $i$ and column $j$ if the branch node with index $j$ is the one above the label node with index $i$ at that particular level (or when no such branch exists, if it is the chosen replacement).
Each row of the matrix has exactly one, and the number each column has is equal to the width of the corresponding branch.

\subsubsection{Level Masks}
For each level matrix there is a corresponding ``mask'', which is a boolean vector that encodes whether each label is on the ``true'' or ``false'' path from that level.
For each label, we look at the corresponding branch above it (the same one determined by the level matrix).
If the label is under the ``true'' path of that branch, we put a $0$ in the corresponding slot of the mask vector; otherwise, we put a $1$.
Thus, given a vector of decision results for the branches above each label, XOR'ing this vector with the ``mask'' yields a new vector which has a $1$ for any label that could be chosen by the decision result at that level.
This means that multiplying (or AND'ing) together each of these vectors would result in a $1$ only for the labels that each tree outputs.

\subsection{Algorithm}
%\milind{If we have space, I think this should be more precisely written up as an algorithm in pseudocode}
The actual inference algorithm is implemented as vectorized computations using these structures. The overall flow is shown in Algorithm~\ref{algo:sever}.
First, the \texttt{SecComp} \cite{blindfold} primitive is applied to Diane's feature vector ({\sf Feats}) and the padded threshold vector from Maurice's model ({\sf Thresh}), which produces a boolean vector of decision results and sentinels.
This vector is multiplied by the reshuffling matrix ({\sf Reshuf}) using  to produce a new boolean vector whose decision results correspond exactly to the branches of the forest in a preorder enumeration.
For each level of the forest, the reshuffled vector is multiplied by the matrix for that level ({\sf Lvls}) and then added to the mask for that level ({\sf Masks}).
Finally, every such vector is multiplied together to produce a single vector with a slot for each leaf node in the forest ({\sf Labels}).
This vector is sent back to Diane for decryption.
By expressing the entire algorithm in terms of these vector operations and matrix multiplications, we are able to exploit a great degree of parallelism, and effectively scale the secure inference process to larger models.

\begin{algorithm}[h!]
    \KwIn{Maurice: {\sf Thresh}, {\sf Reshuf}, {\sf Lvls}, {\sf Masks}}
    \KwIn{Diane: {\sf Feats}}
    Decisions $\gets$ \texttt{SecComp}({\sf Thresh}, {\sf Feats})\;
    Branches $\gets$ \texttt{MatMul}({\sf Reshuf}, {\sf Decisions})\;
    LvlResults $\gets \emptyset$\;
    \ForAll{$i\gets 1$ \textbf{to} NumLevels}{
        LvlDecisions $\gets$ \texttt{MatMul}({\sf Lvls}[i], Decisions)\;
        LvlResults[i] $\gets$ LvlDecisions $\oplus$ {\sf Masks}[i]\;

    }
    {\sf Labels} $\gets$ \texttt{MultAll}(LvlResults)\;
    \KwOut{{\sf Labels}}
    \caption{Algorithm for vectorized inference}
    \label{algo:sever}
\end{algorithm}

Our algorithm uses Aloufi et al.'s SecComp \cite{blindfold} and Shoup's MatMul \cite{helib-algos} as subroutines.
The ability to express most of the computation in terms of MatMul is the key to the algorithm's vectorizability, since the MatMul routine is itself a set of parallel vector operations with constant multiplicative depth.
Performing the computation for each level of the tree at once and then combining them all at the end lets us have a multiplication circuit that is only logarithmic in the forest depth, instead of the naive approach with linear depth.
Section~\ref{sec:complexity} discusses the complexity and multiplicative depth of both the primitives and the algorithm as a whole in more detail.
% \section{Preliminaries}\label{sec:preliminaries}
% \input{preliminaries}
\section{Compiler and Runtime}\label{sec:compiler}
%\raghav{The core idea of this section is taken from the implementation section from before, but with several (stylistic or otherwise) edits.}

While the evaluation algorithm described above is effective at vectorizing the inference of a decision forest, it is not the most natural way in which such models are usually expressed.
A compiler can solve this problem by taking a more natural representation of a trained model and automatically generating a program that creates these vectorizable structures and performs the inference algorithm.
In this section, we discuss the implementation details of such a compiler. \footnote{The latest version of the \system compiler and runtime are available at \reponame}

\paragraph{Input Representation}
The input to the compiler is a serialized trained decision forest model.
The format consists of a line defining the label names as strings, followed by a line for each tree in the forest.
% The trees are serialized in a depth-first left-to-right preorder. \raghav{right word?} \milind{don't you mean something like "in depth-first, left-to-right pre-order"} \raghav{yes I do.}
Each leaf node outputs the index of the label it corresponds to.
For every branch node, the serialized output contains the index of its feature, the threshold value its compared to, and the serializations of its left and right subtrees respectively.

\paragraph{Compiler Architecture}
\system is a staging metacompilation framework.
The input to the first stage is a serialized decision forest model.
The \system compiler translates this to a C++ program that uses the vectorizable data structures described in Section~\ref{sec:primitives}, specialized to the given model, and invokes the algorithmic primitives provided by the \system runtime.
The generated C++ program is then compiled and linked against the \system runtime library to produce a binary which can be executed to perform secure inference queries.

Structuring \system as a staging compiler allows us to specialize the generated C++ code by (1) choosing an appropriate set of encryption parameters for the model being compiled and (2) selecting optimal implementations for the algorithmic primitives given the FHE protocol and implementation used by the runtime.
In our sensitivity analysis in Section~\ref{sec:evaluation}, we performed a sweep over the possible encryption parameters and found that for the models we were compiling, a single set dominated all the others.
Since \system is currently targeted only to use the BGV implementation in HElib, a single set of optimal implementations is used for all the primitives.
However, if \system were to use a different protocol and backend (for instance, SEAL and CKKS), these choices could matter and the staging compiler could appropriately tune the parameters and implementations.

\paragraph{\system Runtime}
The runtime has datatypes that represent both plaintext and ciphertext vectors and matrices, as well as the parties playing the role of {\em model owner} (Maurice), {\em data owner} (Diane), and {\em evaluator} (Sally).
It also exposes primitives to encrypt and decrypt models and feature vectors, and securely execute an inference query given an encrypted model and encrypted feature vector.
The programmer can use these datatypes to encode their application logic, and then link against the generated C++ code to produce a binary that securely performs decision forest inference.

We use the HElib library \cite{HElib} with the BGV protocol \cite{BGV} as our framework for homomorphic encryption.
This library provides low-level primitives for encrypting and decrypting plaintext and ciphertexts, homomorphically adding and multiplying ciphertexts, and generating public/secret key pairs.
HElib also supports ciphertext packing which gives us the vectorizing capabilities we need.
\section{Complexity Analysis}
\label{sec:complexity}

This section characterizes the complexity of \system.
This complexity is parameterized on various parameters of the decision forest model: the number of branches $b$, the total number of levels $d$, the fixedpoint precision $p$, and the quantized width $q$.
(Definitions of these parameters can be found in Section~\ref{sec:definitions}.)
The complexity of FHE circuits is characterized by two elements: (1) the number of each kind of primitive FHE operation and (2) the {\em multiplicative depth} of the FHE circuit.
The former captures the ``work'' needed to execute the circuit.
The latter, characterized by the longest dependence chain of multiplications in the circuit, determines the encryption parameters needed to evaluate the circuit accurately (higher multiplicative depth requires more expensive encryption, or bootstrapping).

The FHE operations used to express the amount of work are: (1) {\em Encrypt}, which produces a single ciphertext from a plaintext bitvector; (2) {\em Rotate}, which rotates all the entries in a vector by a constant number of slots; (3) {\em Add}, which computes the XOR of two encrypted bitvectors; (4) {\em Multiply}, which computes the AND of two encrypted bitvectors, and (5) {\em Constant Add}, which computes the XOR of an encrypted bitvector with a plaintext one.
The {\em Multiply} operation incurs a multiplicative depth of 1, and all the rest incur a multiplicative depth of 0.

Table~\ref{tab:algorithm-step-costs} characterizes the steps of the \system algorithm, in terms of the number of FHE operations and their multiplicative depth, as well as the cost of encrypting the data and models (which do not factor in to multiplicative depth, as they are separate from the circuit).
Table~\ref{tab:complexities} shows the overall cost of \system, including combining the multiplicative depths of the individual steps according to their dependences in the overall circuit.
Note that the cost of processing a single level is incurred $d$ times, but the level processing steps occur in parallel in the FHE circuit, so altogether the level processing only contributes $1$ to the multiplicative depth.

% Table~\ref{tab:operation-descriptions} shows the operations used.
% \begin{table}
%     \centering\tablesize
%     \caption{Descriptions of operations used in protocol}
%     \label{tab:operation-descriptions}
%     \begin{tabular}{ll}
%         \toprule
%         Operation    & Description                                                              \\
%         \midrule
%         Encrypt      & Encrypt a vector of plaintext bits into a single ciphertext              \\
%         Rotate       & Rotate all entries in a vector by a constant number of slots             \\
%         Add          & Add two ciphertxt vectors (bitwise XOR)                                  \\
%         Constant Add & Add a plaintext bit to each element of a ciphertext vector (bitwise NOT) \\
%         Multiply     & Multiply two ciphertext vectors (bitwise AND)                            \\
%         \bottomrule
%     \end{tabular}
% \end{table}
\begin{table}

    \caption{Operation counts and multiplicative depth for \system}
    \label{tab:algorithm-step-costs}

    \vspace{-1em}

    \begin{subtable}{0.48\textwidth}
        \centering\tablesize
        \caption{Complexity for Secure Comparison}
        \begin{tabular}{ll}
            \toprule
            Operation    & Number of Ops       \\
            \midrule
            % Encrypt & $0$\\
            % Rotate & $0$\\
            Add          & $4 p-2$             \\
            Constant Add & $p$                 \\
            Multiply     & $p\log p + 3 p - 2$ \\
            \bottomrule
        \end{tabular}

        \vspace{0.5em}
        Multiplicative depth:  $2\log p + 1$
        \vspace{0.75em}
    \end{subtable}

    %    \begin{subtable}{0.48\textwidth}
    %        \centering\tablesize
    %        \caption{Complexity for $m\times n$ Matrix Multiplication}
    %        \begin{tabular}{ll}
    %            \toprule
    %            Operation & Description \\
    %            \midrule
    %            % Encrypt & $0$\\
    %            Rotate    & $n$         \\
    %            Add       & $n$         \\
    %            % Constant Add & $0$\\
    %            Multiply  & $n$         \\
    %            Depth     & $1$         \\
    %            \bottomrule
    %        \end{tabular}
    %    \end{subtable}

    \begin{subtable}{0.48\textwidth}
        \centering\tablesize
        \caption{Complexity for processing a single level (repeats $d$ times)}
        \begin{tabular}{ll}
            \toprule
            Operation & Number of Ops \\
            \midrule
            % Encrypt & $0$\\
            Rotate    & $b$           \\
            Add       & $b+1$         \\
            % Constant Add & $0$\\
            Multiply  & $b$           \\
            \bottomrule
        \end{tabular}

        \vspace{0.5em}
        Multiplicative depth: $1$
        \vspace{0.75em}
    \end{subtable}

    \begin{subtable}{0.48\textwidth}
        \centering\tablesize
        \caption{Complexity for accumulating results from all levels}
        \begin{tabular}{ll}
            \toprule
            Operation & Number of Ops \\
            \midrule
            % Encrypt & $0$\\
            % Rotate & $0$\\
            % Add & $0$ \\
            % Constant Add & $0$\\
            Multiply  & $2d-2$        \\
            \bottomrule
        \end{tabular}

        \vspace{0.5em}
        Multiplicative Depth: $\log d$
        \vspace{0.75em}
    \end{subtable}

    \begin{subtable}{0.48\textwidth}
        \centering\tablesize
        \caption{Complexity for encrypting model}
        \begin{tabular}{ll}
            \toprule
            Operation & Number of Ops \\
            \midrule
            Encrypt   & $p+q+d(b+1)$  \\
            % Rotate & $0$\\
            % Add & $0$ \\
            % Constant Add & $0$\\
            % Multiply & $0$ \\
            % Depth & $0$\\
            \bottomrule
        \end{tabular}
        \vspace{0.75em}
    \end{subtable}

    \begin{subtable}{0.48\textwidth}
        \centering\tablesize
        \caption{Complexity for encrypting data}
        \begin{tabular}{ll}
            \toprule
            Operation & Number of Ops \\
            \midrule
            Encrypt   & $1$           \\
            % Rotate & $0$\\
            % Add & $0$ \\
            % Constant Add & $0$\\
            % Multiply & $0$ \\
            % Depth & $0$\\
            \bottomrule
        \end{tabular}
    \end{subtable}
\end{table}

\begin{table}
    \centering\tablesize
    \caption{Total Evaluation Complexity}
    \vspace{-1em}
    \begin{tabular}{ll}
        \toprule
        Operation    & Number of Ops                       \\
        \midrule
        Encrypt      & $1+p+q+d(b+1)$                      \\
        Rotate       & $q+d b$                             \\
        Add          & $4 p - 2 + q + d  (b + 1)$          \\
        Constant Add & $p$                                 \\
        Multiply     & $p\log p + 3 p + q + d b + 2 d - 4$ \\
        \bottomrule
    \end{tabular}

    \vspace{0.5em}
    Multiplicative Depth: $2\log p + \log d + 2$
    \label{tab:complexities}
\end{table}
\section{Security Properties}\label{sec:parties}
This section describes the various security properties of decision forest programs built using \system. Section~\ref{sec:app-security} discusses information leakage between the parties, while Section~\ref{sec:app-security-label-format} discusses the privacy implications of different {\em design} decisions in \system.
%

%Section~\ref{sec:app-security} discusses the different ways the parties (Diane, Maurice, and Sally) can be configured in a two-party FHE system like HElib, and describes what information leaks between the parties in each configuration.
%%
%Section~\ref{sec:app-security-label-format} discusses the privacy implications of different {\em design} decisions we made.
% \subsection{Party Setup}
% \input{security}
\subsection{Information Leakage}
\label{sec:app-security}

%\begin{table}[t]
%    \centering\tablesize
%    \caption{Values owned by each party}
%    \label{tab:data-ownership}
%    \begin{tabular}{lc}
%        \toprule
%        Party & Values                                                   \\
%        \midrule
%        S     & $\emptyset$                                              \\
%        M     & $\tau$, $\mathscr{L}$, $\mathscr{R}$, $m$, $q$, $b$, $K$ \\
%        D     & $f$                                                      \\
%        \bottomrule
%    \end{tabular}
%\end{table}
\begin{table*}[t]
    \centering\tablesize
    \caption{Data revealed to each notional party in two-party configurations}
    \label{tab:data-revealed}
    \begin{tabular}{lccc}
        \toprule
        Scenario   & Revealed to $S$    & Revealed to $M$ & Revealed to $D$ \\
        \midrule
        $S, M = D$ & $q$, $b$, $d$      & $\emptyset$     & $\emptyset$     \\
        $S = M, D$ & $\emptyset$        & $\emptyset$     & $K$, $b$        \\
        $S = D, M$ & $q$, $b$, $K$, $d$ & $\emptyset$     & $q$, $b$, $K$   \\
        \bottomrule
    \end{tabular}
\end{table*}
\begin{table*}[t]
    \centering\tablesize
    \caption{Data revealed to each party in three-party configurations}
    \label{tab:data-revealed-3p}
    \begin{tabular}{lccc}
        \toprule
        Scenario                     & Revealed to $S$    & Revealed to $M$ & Revealed to $D$ \\
        \midrule
        $S, M, D$, no collusion      & $q$, $b$, $d$, $K$ & $\emptyset$     & $K$, $b$        \\
        $S, M, D$, S colludes with M & everything         & everything      & $K$, $b$        \\
        $S, M, D$, S colludes with D & everything         & $\emptyset$     & everything      \\
        \bottomrule
    \end{tabular}
\end{table*}

\system has three {\em notional} parties: the model owner {\bf M}aurice, the data owner {\bf D}iane, and the server {\bf S}ally. Maurice owns $\tau$, $\mathscr{L}$, $\mathscr{R}$, $m$, $q$, $b$, and $K$, while Diane owns the feature vector $f$. Sally owns nothing.

\paragraph{Two Physical Parties}
FHE is inherently a two-party protocol, so although the secure inference problem has three notional parties, our system focuses on the cases where there are only two \emph{physical} parties (i.e.,  two of the notional parties are actually the same person).
There are three scenarios:
\begin{enumerate}
    \item Where $M = D$; for instance, if the model and data are owned by the same party, which offloads the inference to an untrusted server. This is the standard ``computation offloading'' model used by most FHE applications \cite{CHET,Alchemy,EVA}.
    \item Where $M = S$; if the model is stored on some server which allows clients to send encrypted data for classification
    \item Where $D = S$; if the model is trained and sent directly to a client for inference, but the client must be prevented from reverse-engineering the model.
\end{enumerate}

In Table~\ref{tab:data-revealed} we describe what data is explicitly revealed or implicitly leaked to each party.
When $M = D$, obviously neither party can leak information to the other.
However, because matrices are encrypted as a vector of ciphertexts with one per column (diagonal), $S$ learns the number of columns in each matrix.
This translates to learning the number of branches $b$ from each level matrix $\mathscr{L}$, and learning the quantized width $q$ from the reshaping matrix $\mathscr{R}$.
Furthermore, since the level masks and matrices are stored separately, $S$ also learns the maximum depth of the forest.

When $S = M$, neither $S$ nor $M$ can leak information to each other.
However, $M$ must explicitly send the value of $K$ to $D$ to get feature vectors with the right padding.
When the inference result is sent back, $D$ also learns $b + 1$, as it is the length of the final inference vector.

When $S = D$, once again neither $S$ nor $D$ leak information to each other.
However, this time $M$ not only reveals $K$ and $b$ to both $S$ and $D$ the same way as in case (2), but $q$ is also leaked through the widths of the matrices, as well as $d$.

\paragraph{Three Parties}
When $S$, $M$, and $D$ are separate physical parties that do not collude, $M$ necessarily leaks to $S$ the values of $b$, $q$, and $d$, as well as revealing $K$.
$S$ then reveals $K$ to $D$, and $M$ leaks $b$ to $D$.
Even though $M$ and $D$ use the same key pair, because neither colludes with $S$, neither ever gets access to the other's ciphertexts, and privacy between the two is therefore preserved.
However, if one of the parties does collude with $S$, they gain access to the other party's ciperhtexts which can then be easily decrypted.
Thus in the case where there is collusion between $M$ or $D$ and $S$, everything is leaked.
Table~\ref{tab:data-revealed-3p} summarizes these results.

Since it is difficult to convince both $M$ and $D$ that the other is not colluding with $S$, we see that attempting to run this protocol with three physical parties using single-key FHE is unreasonable.
There has been a lot of prior work on multikey FHE schemes \cite{EfficientMKFHE,chen2017BatchedMM} and threshold FHE, which uses secret sharing to extend single-key FHE to work in a multiparty setting~\cite{asharov}. These schemes act as ``wrappers'' that construct a new, joint key pair for FHE (in this case shared by $D$ and $M$), and hence can be applied directly to \system at the cost of introducing additional rounds of communication and additional encryption/decryption steps.
\subsection{Security implications of \system design}\label{sec:app-security-label-format}

The design of \system admits different points in the design space that trade off security and performance. Here, we discuss the implications of the design points that we chose.

\subsubsection{Feature padding}

Choosing to have Diane replicate and pad her feature vector is a tradeoff we make between the performance and security of \system.
To avoid requiring Diane to do any preprocessing beyond replicating her feature vector, we would need to explicitly reveal the multiplicity of each feature used in the model.
By requiring the feature vector to be padded, we only reveal the maximum feature multiplicity of the model must be explicitly revealed.

We could even avoid revealing the exact maximum multiplicity, and instead only reveal an upper bound, simply by adding several extra sentinel values to each feature in the threshold vector.
The performance overhead of this would be minimal, except a slightly more expensive matrix multiply to remove the extra sentinel values (the size of this overhead scales with how loose the given upper bound is).

To avoid leaking {\em any} multiplicity information, we could also relax the requirement that the Diane replicate her features at all, instead accepting a vector that lists each feature once, and requiring that the server carry out the necessary replication directly on the ciphertext vector.
While this does prevent Diane from learning anything about feature multiplicities in the mode, it has the effect of replacing several (cheap) plaintext replication operations with their equivalent ciphertext ones, which are much more expensive.
%\raghav{How's that?}

\subsubsection{Returning classification bitvectors}
%\milind{Fold this into the above}\raghav{One day I'll structure an entire paper as a single catamorphism.}
Returning the bitvector of classification results rather than accumulating them to return a single label leaks some information about the structure of the model to Diane.

First, it requires a ``codebook'' (i.e. a map from each position in the bitvector to the label it represents) to be revealed to Diane. This reveals the order of the labels in the constituent trees of the forest (though not the ``boundaries'' between the trees). It is possible to avoid leaking the order of the label nodes by first having the server generate a random permutation to apply to the decision result bitvector (via a plaintext matrix/ciphertext vector multiplication), then applying the same permutation to the codebook.

%\milind{This sentence below is unclear: it makes it sound like you're talking about feature multiplicity, rather than classification result multiplicity} \raghav{Better?}
% However, this still reveals how many times each label appears in a leaf node of the model (i.e. whether a particular classification result appears more frequently than another).
Shuffling the codebook still reveals information about the model structure; in particular, it leaks how many leaf nodes correspond to each label. For instance, knowing whether a particular label is output by most of the leaves in the forest versus only being output by a single leaf potentially reveals something about how ``likely'' that label is to be chosen.
This can also be avoided if the server pads both the codebook and the classification result bitvector with random extra labels before returning both of them; this step can be folded into the shuffling step as well, so it has minimal extra cost.

\system currently assumes that the codebook is already known to Diane, and performs neither shuffling nor padding.

%\milind{Is this unavoidable? Or is it only unavoidable because we chose to make a tradeoff? ;-)}\raghav{Fixed it :P}
Another data leak is the label result chosen by each tree.
In other words, Diane learns that, for instance, two trees chose $X$ and three chose $Y$, rather than simply learning that the final classification is $Y$.
This is an unavoidable consequence of \system's design of returning a bitvector of results rather than performing the reduction server-side. The effect of this design is to put the burden of accumulating all the chosen labels into a single classification on Diane. Doing so necessarily requires revealing all the chosen labels.

Accumulation could be done by Sally to avoid this leak, at the cost of expensive ciphertext operations, including potentially multiple interactive rounds to change the plaintext modulus and count up the occurrences of each label.
Thus, while this design is somewhat more secure, it adds {\em communication complexity} in addition to computational complexity.

% Since there is no collusion, even though the fact that $M$ and $D$ use the same key pair 
% Here we discuss the security characteristics of the algorithm.
% what data is hidden
% The fact that we can tune how much is hidden by doing matrices in plaintext
% not safe against collusion between D/M and S, but can be extended to multikey FHE

% \section{Implementation}\label{sec:implementation}
% \input{implementation}

\section{Evaluation}\label{sec:evaluation}
We evaluate \system in several ways.
First, we evaluate how well \system performs against the prior state-of-the-art in secure decision forest inference, \citet{blindfold}.
This evaluation looks at sequential and parallel performance, and focuses on the classic, offloading-focused privacy model where the model and data are owned by one party, and the server is another party (see Section~\ref{sec:parties}).
Second, we consider \system's ability to handle different party configurations, in particular, when the server and model are owned by one party, and the data by another.
Finally, we use microbenchmarks to understand \system's sensitivity to different aspects of the models: depth, number of branches, and feature precision.

\subsection{Benchmarks, Configurations, and Systems}
\label{sec:evaluation-configurations}

To evaluate \system, we synthesized several microbenchmark models that varied the number of levels, number of branches, and the bits of precision used for expressing thresholds.
We use these microbenchmarks both for performance studies (this section) and for sensitivity studies (Section~\ref{sec:compiler}. %sec:microbenchmarks
In addition to microbenchmarks, we obtained open-source ML data sets to train decision forests for real-world benchmarks, {\tt income}~\cite{income} and {\tt soccer}~\cite{soccer}.
We used the scikit-learn library \cite{scikit-learn} to train random forest classifiers on these data sets.
For each of the real-world data sets, we generated two differently-sized models (suffixed {\tt 5} and {\tt 15}), reflecting the number of decision trees comprising each forest.

Configuring HELib involves setting several encryption parameters: the {\em security parameter}, the number of {\em bits} in the modulus chain, and the number of {\em columns} in the key-switching matrices.
Increasing the security parameter results in larger ciphertexts, increasing security and computation time; increasing the number of bits in the modulus chain increases the maximum multiplicative depth the circuit can reach; and changing the number of columns in the key-switching matrices affects the available vector widths.
We performed a sweep over the range of possible encryption parameters for our models, and found a single set of parameters that worked sufficiently well.
Table~\ref{tab:optimal-params} lists the encryption parameters we used.
(Note that it is possible that for other models, or other FHE implementations, other parameters will be superior; autotuning these parameters can be incorporated into the staging process, as described in Section~\ref{sec:compiler}.)

All experiments were performed on a 32-core, 2.7 GHz Intel Xeon E5-4650 server with 192 GB of RAM. Each core has 256 KB of L2 cache, and each set of 8 cores shares a 20 MB last-level-cache.

\begin{table}
    \centering\tablesize
    \caption{Optimal encryption parameter values}
    \vspace{-1em}
    \begin{tabular}{lr}
        \toprule
        Parameter          & Value \\
        \midrule
        Security Parameter & 128   \\
        Bits               & 400   \\
        Columns            & 3     \\
        \bottomrule
    \end{tabular}
    \vspace{-1em}
    \label{tab:optimal-params}
\end{table}

\subsection{\system Performance}
\label{sec:evaluation-performance}

Our first evaluation focuses on the sequential and parallel performance of \system.
Our baseline for \system is the state-of-the-art approach for performing secure decision forest evaluation in FHE, by~\citet{blindfold}.
Because Aloufi et al.'s implementation was not available, we implemented their algorithms ourselves.
We made our best effort to optimize our reimplementation, including introducing parallelism with Intel's Thread Building Blocks~\cite{tbb} (indeed, our implementation appears to scale better than Aloufi et al.'s reported scalability).
Crucially, both our baseline and \system use the same FHE library, and the same implementation of SecComp, which was introduced by \citet{blindfold}.

We evaluated both implementations on two primary criteria: how quickly the compiled models could execute inference queries, and how effectively \system was able to take advantage of parallelism to scale to larger models.
For each model, we performed 27 inference queries, in both single-threaded and multithreaded mode. We report the median running time across these queries (confidence intervals in all cases were negligible).

\begin{figure}
    \centering
    \includegraphics[width=0.95\columnwidth]{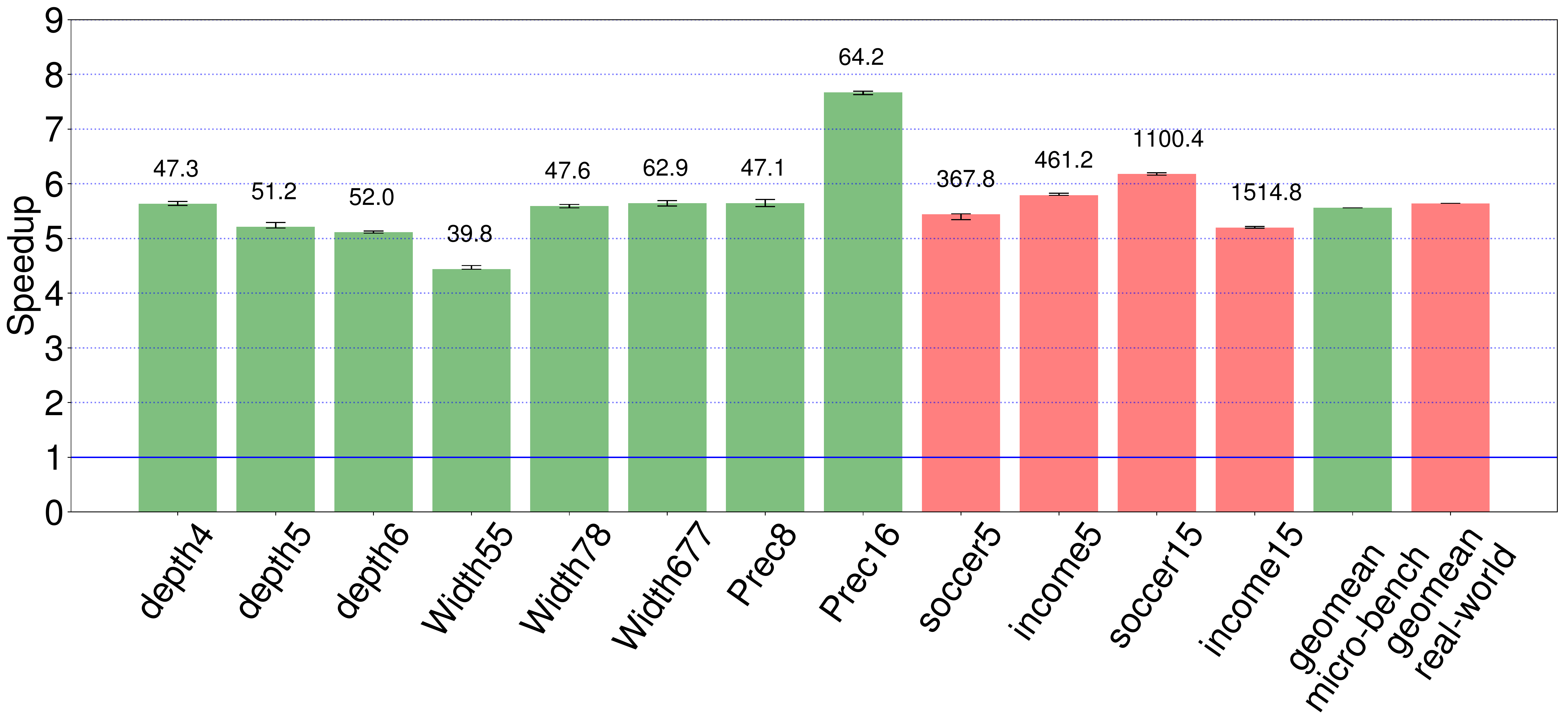}
    \vspace{-1em}
    \caption{Speedup of \system-compiled models over our implementation of \citet{blindfold} when both are single-threaded. The number on top of each bar is the median running time (in milliseconds) for that model using \system.}
    \Description{Bar graph depicting the speedup for each model when compiled using \system, using \citet{blindfold} as a baseline}
    \label{fig:single-threaded-speedup}
\end{figure}

%\milind{The labels in these figures are still unreadably small...}
Figure~\ref{fig:single-threaded-speedup} shows the relative speedups over prior work for each model compiled using \system.
%
% \milind{These numbers need to be updated!}
We see that we have a substantial speedup over the baseline, ranging from 5$\times$ to over 7$\times$, with a geometric mean of close to $6\times$.

\paragraph{Multithreading}
Next, we ran inference queries on the models with multithreading enabled.
For all queries, we ran the systems using 32 threads.
While the individual queries were mutithreaded, they were still executed sequentially one after another.

\begin{figure}
    \centering
    \includegraphics[width=0.95\columnwidth]{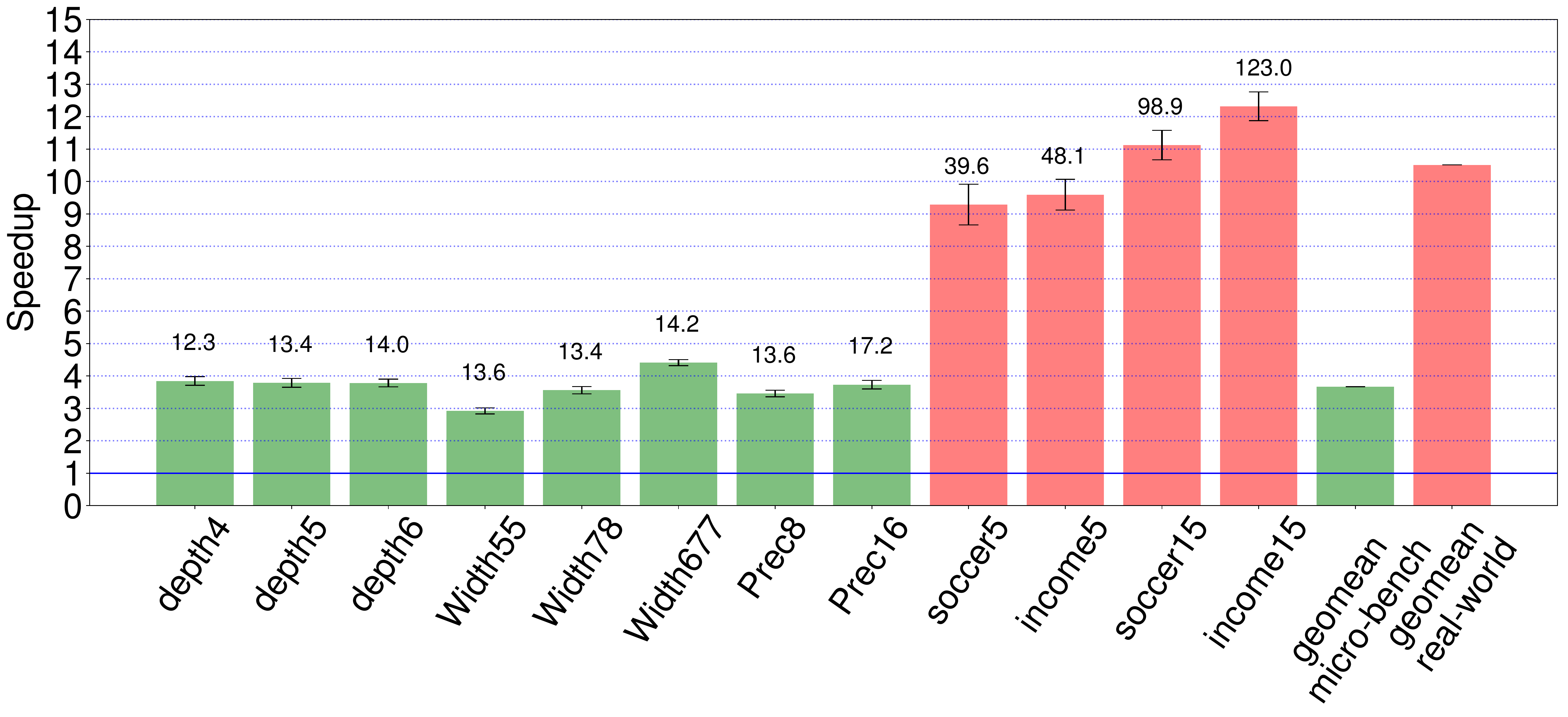}
    \vspace{-1em}
    \caption{Speedup that \system-compiled models experience when multithreaded instead of single-threaded. The number on top of each bar is the median run-time (in milliseconds) for multithreaded inference.}
    \Description{Bar graph depicting the speedup for each model when multithreading inference queries}
    \label{fig:mt-over-st-speedup}
\end{figure}

Figure~\ref{fig:mt-over-st-speedup} shows the multithreaded speedup of \system over single-threaded \system.
Note that in this study, we evaluated even larger models for our real-world datasets, as \system is able to evaluate these in a reasonable amount of time while our baseline is not.
We see that while parallel speedup for the microbenchmarks is relatively modest (around 2.5$\times$), parallel speedup for the real-world models is much better (almost $5\times$), as we might expect: the real-world models are larger, and present more parallel work.

\begin{figure}
    \centering
    \includegraphics[width=0.95\columnwidth]{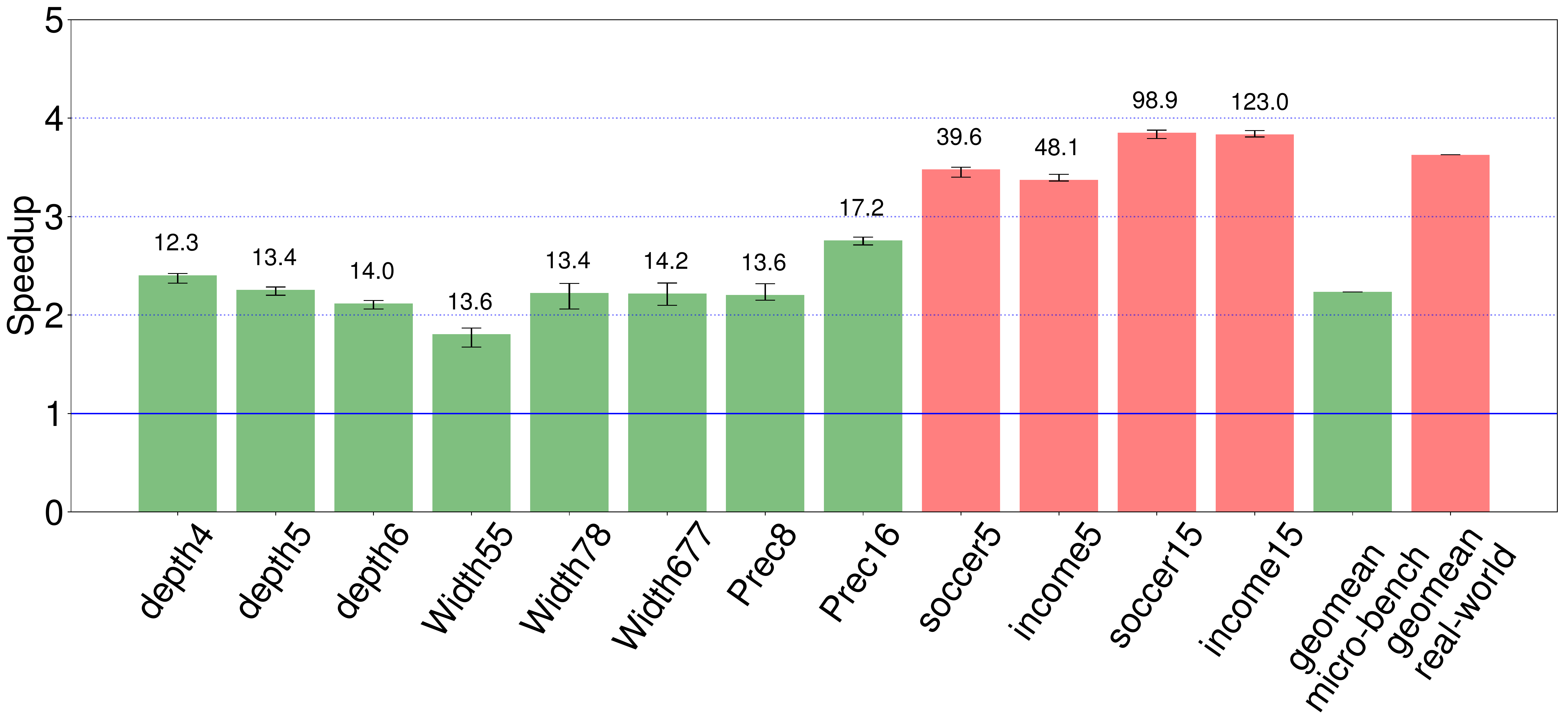}
    \vspace{-1em}
    \caption{Speedup of \system-compiled models over our implementation of \citet{blindfold} when both are multithreaded.  The number on top of each bar is the median run-time (in milliseconds) for multithreaded \system.}
    \Description{Bar graph depicting the speedup of \system over Aloufi, et al. when executing with multithreading enabled. The number on top of each bar is the median run-time (in milliseconds) for multithreaded \system.}
    \label{fig:multi-threaded-speedup}
\end{figure}

Figure~\ref{fig:multi-threaded-speedup} compares \system's parallel performance to our baseline.
We note that it appears that \system scales worse than the baseline (if they scaled equally well, the speedups in Figure~\ref{fig:multi-threaded-speedup} would match the speedups in Figure~\ref{fig:single-threaded-speedup}).
The source of this seeming shortcoming is subtle.
\system's design makes heavy use of ciphertext packing to amortize the overheads of FHE computation.
This ciphertext packing essentially consumes some of the parallel work in the form of ``vectorized'' operations, even in the single-threaded case, leaving the \system implementations with less parallelism to exploit using ``normal'' parallelization techniques.
In contrast, all of these parallelism opportunities can {\em only} be exploited by multithreading in the baseline (and recall that our baseline implementation appears to scale better than the original implementation).
We can observe this effect by noting that the gap in scaling is smaller for the larger, real-world models ({\tt income5} and {\tt soccer5}): there is more parallelism to start with, so even after ciphertext packing, there is more parallelism for \system to exploit.

\subsection{Different Party Setups}
\begin{figure}
    \centering
    \includegraphics[width=0.95\columnwidth]{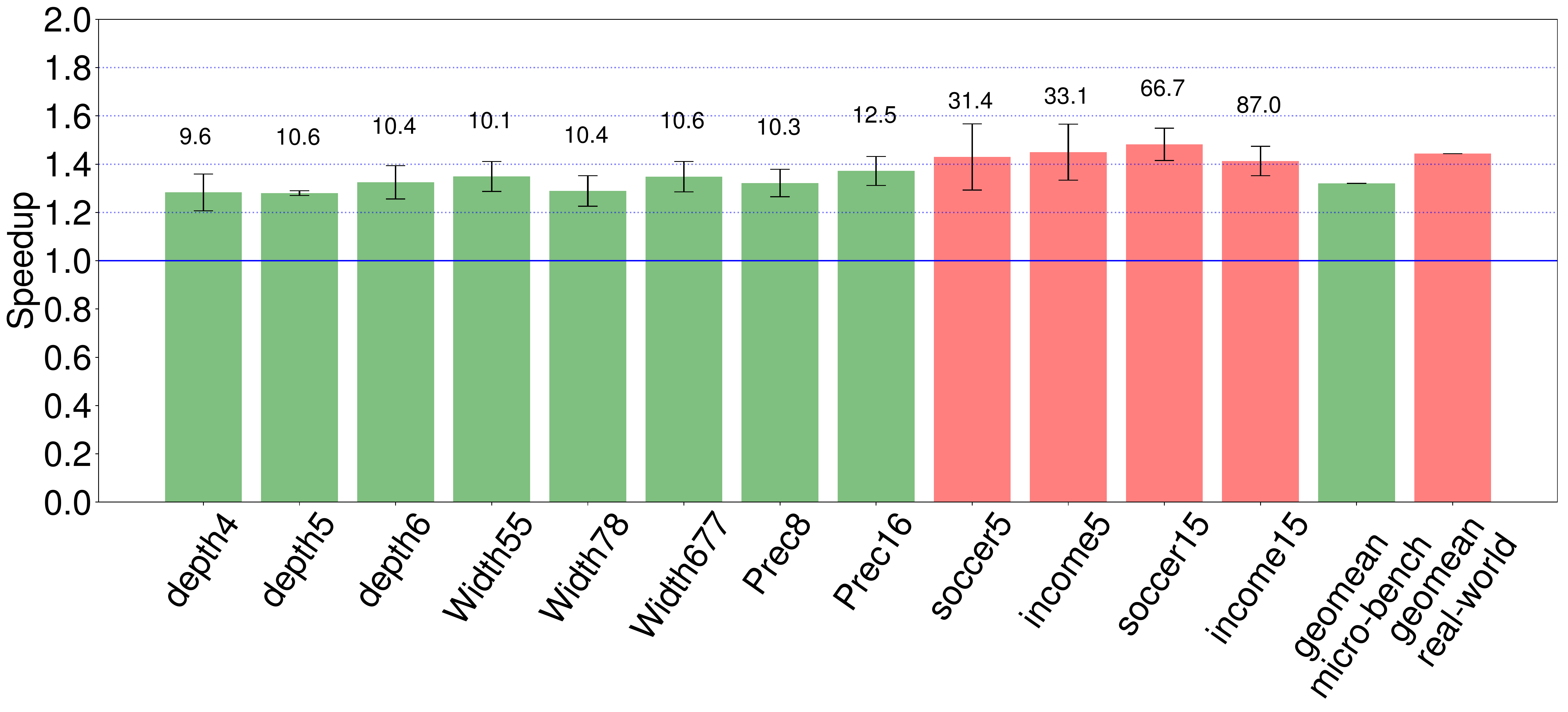}
    \vspace{-1em}
    \caption{Speedup of inference queries executed on plaintext models (when Maurice = Sally) compared to encrypted models (when Diane = Maurice). The number on top of each bar show the median inference run-time (in milliseconds) on the plaintext models.}
    \Description{Bar graph depicting speedup from evaluating models in plaintext}
    \label{fig:ptxt-mod-speedup}
\end{figure}

As discussed in Section~\ref{sec:parties}, the two-party setup used by \system admits different configurations for the identities of these parties. We would expect to see speedup if Maurice and Sally are the same party (so the models can be represented in plaintext) compared to when Maurice and Diane are the same party (o the model has to be encrypted).
Figure~\ref{fig:ptxt-mod-speedup} shows the speedup of inference when using the second configuration (plaintext models) versus the first (ciphertext models).
As expected, we see that plaintext models result in substantial speedups of roughly 1.4$\times$.

\subsection{Evaluation on Microbenchmarks}

\begin{figure*}
    \begin{subfigure}{0.33\linewidth}
        \centering
        \Description{Stacked bar graph showing run time of depth 4, 6, and 8 microbenchmarks}
        \includegraphics[width=1.1\linewidth]{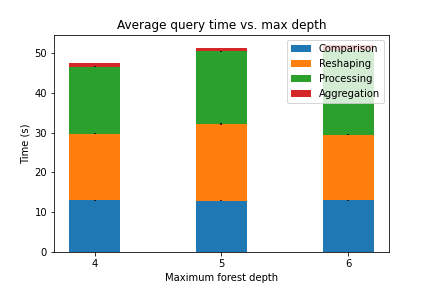}
        \vspace{-2em}
        \caption{Run time vs. max depth}
        \vspace{-0.5em}
        \label{fig:micro_depth}
    \end{subfigure}
    \begin{subfigure}{0.33\linewidth}
        \centering
        \Description{Stacked bar graph showing run time of 10, 15, and 20 branch microbenchmarks}
        \includegraphics[width=1.1\linewidth]{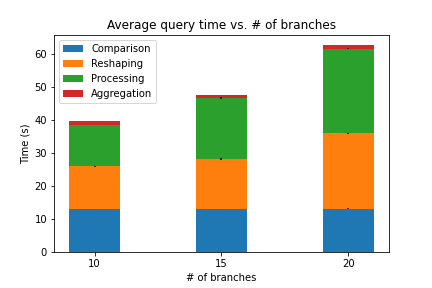}
        \vspace{-2em}
        \caption{Run time vs. branches}
        \vspace{-0.5em}
        \label{fig:micro_branches}
    \end{subfigure}
    \begin{subfigure}{0.33\linewidth}
        \centering
        \Description{Stacked bar graph showing run time of 8- and 16-bit microbenchmarks}
        \includegraphics[width=1.1\linewidth]{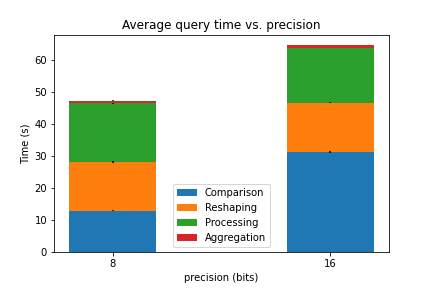}
        \vspace{-2em}
        \caption{Run time vs. precision}
        \label{fig:micro_precision}
    \end{subfigure}
    \vspace{-1em}
    \caption{Run time of microbenchmarks}
    \Description{Stacked bar graph showing run time of each microbenchmark}
    \label{fig:microbenchmarks}
\end{figure*}

\begin{table}[t]
    \footnotesize
    \centering\tablesize
    \caption{Microbenchmark specifications}
    \vspace{-1em}
    \begin{tabular}{lcccc}
        \toprule
        Model name & Max. depth & Precision & \# of trees & \# q of branches \\
        \midrule
        depth4     & 4          & 8         & 2           & 15               \\
        depth5     & 5          & 8         & 2           & 15               \\
        depth6     & 6          & 8         & 2           & 15               \\
        width55    & 5          & 8         & 2           & 10               \\
        width78    & 5          & 8         & 2           & 15               \\
        width677   & 5          & 8         & 3           & 20               \\
        prec8      & 5          & 8         & 2           & 15               \\
        prec16     & 5          & 16        & 2           & 15               \\
        \bottomrule
    \end{tabular}
    \label{tab:micro-specs}
\end{table}

To better understand the different components of \system, we used eight randomly-generated forests with different properties: feature precision, maximum levels, number of trees, and number of branches.
The details on the size of each forest can be found in Table~\ref{tab:micro-specs}.
Every forest had 2 features and 3 distinct labels.
Figure~\ref{fig:microbenchmarks} shows the median running time for each model, broken down by the time each step took (comparison, reshuffling, processing levels, and aggregating).
To facilitate comparison, each sub-figure compares models that are similar except for a particular parameter under test.

\paragraph{Effects of depth}
Figure~\ref{fig:micro_depth} shows models that differ in terms of tree depth.
Comparison and reshaping times are largely unaffected by the maximum forest depth, whereas the total level processing time increases approximately linearly.
Aggregation time is logarithmic in depth, but this is negligibly small compared to the rest of the evaluation.
This makes sense because at each depth level there is approximately an equal amount of work to be done.

\paragraph{Effects of branching}
Figure~\ref{fig:micro_branches} shows models that differ in terms of number of branches.
While the comparison time is unaffected by branching, both the reshaping time and total level processing time are.
The relationship between branching and reshaping is close to linear, as reshaping actually depends linearly on the quantized branching.
By contrast, the total level processing time is directly proportional to the number of branches.
This is because the matrix at each level has a number of columns equal to the number of branches.
Thus for a model with twice as many branches, the corresponding matrices will be twice as wide and take twice as long to process without multithreading.

\paragraph{Effects of precision}
Finally, Figure~\ref{fig:micro_precision} shows models that differ in terms of feature precision.
Reshaping, level processing, and aggregation are, as expected, unaffected by changing the model precision.
However, as suggested by the complexity analysis in Section~\ref{sec:complexity}, comparison time increases super-linearly with precision. %\milind{If we move complexity stuff to an appendix, change this to talk about appendices instead.}
This superlinear relationship is evident from the data in Figure~\ref{fig:micro_precision}.

\section{Conclusion}\label{sec:conclusions}
In this paper we presented \system, a staging compiler and runtime for decision forest models that takes advantage of the noninterference semantics of fully-homomorphic encryption to relax control-flow dependences and map these models to vectorizable, FHE-based primitives. \system represents the first approach to exploit ciphertext packing to perform secure, decision forest inference.

We showed that \system can automatically stage decision-forest models, including large-scale ones trained on real-world datasets, into efficient, specialized C++ programs implemented using our novel primitives. We further showed that these implementations significantly outperform state-of-the-art secure inference algorithms, while still scaling well (especially for larger models).

This paper thus represents the next step in building efficient, secure inference procedures for machine-learned models. Future work includes implementing \system's primitives not in terms of low-level FHE libraries like HELib but instead in terms of higher-level FHE-based intermediate languages, like EVA~\cite{EVA}, allowing for further tuning and optimization.

%We tested \system on randomly generated synthetic microbenchmarks to investigate how various parameters such as model precision, maximum levels, and total number of branches affect the inference time of the vectorized models.
%We found that while execution time scales linearly with model size, some of this work (such as the threshold comparison step) is batched into fixed-cost vector operations, \raghav{and much of the remaining work is independent and offers opportunities for multithreading.}

%We also trained benchmark models using open source machine learning data to investigate how well \system scales to the much larger real-world model sizes.
%We compared the inference query times from our vectorized models against our best-faith implementation of the prior state-of-the-art work by \citet{blindfold}, and found that not only does \system compare very favorably in terms of absolute run time, but that we were able to take more advantage of parallelism with our restructuring.
%
%Finally, we investigated the ability of \system to generate FHE code for different party configurations, and found that we could take advantage of situations where the model could be expressed in plaintext to produce programs that executed faster than their ciphertext counterparts.
%\raghav{It still feels like there should be a sentence to tie it all off at the end\dots}

\section*{Acknowledgments}

The authors would like to thank Hemanta Maji for discussions of this problem that inspired the design of the \system. The authors appreciate the feedback from the anonymous reviewers from OOPSLA 2021 and PLDI 2021 that have improved the paper. We would especially like to thank our shepherd, Madan Musuvathi, for his efforts.

This work was partially supported by NSF Grants CCF-1919197 and CCF-1725672. This work was also partially supported by the Office of the Director of National Intelligence (ODNI), Intelligence Advanced Research Projects Activity (IARPA), contract \#2019-19020700004. The views and conclusions contained herein are those of the authors and should not be interpreted as necessarily representing the official policies, either expressed or implied, of ODNI, IARPA, or the U.S. Government. The U.S. Government is authorized to reproduce and distribute reprints for governmental purposes notwithstanding any copyright annotation therein.
\balance
\bibliographystyle{ACM-Reference-Format}
\bibliography{papers}
\pagebreak
\clearpage
\appendix

\pagebreak
%\section{Pseudocode}
%Coming soon, to a PLDI submission near you!
\end{document}